\documentclass[draft]{agujournal2019}
\usepackage{url} 
\usepackage{lineno}
\usepackage[inline]{trackchanges} 
\usepackage{soul}
\usepackage{booktabs}

\usepackage{lscape}
\usepackage{graphicx}
\usepackage{amsmath}
\DeclareMathOperator{\erf}{erf}

\draftfalse

\journalname{JGR: Atmospheres}

\begin{document}

\title{Detecting Extreme Temperature Events Using Gaussian Mixture Models}

\authors{Aytaç Paçal\affil{1,2}, Birgit Hassler\affil{1}, Katja Weigel\affil{2,1}, M. Levent Kurnaz \affil{4}, Michael F. Wehner\affil{3}, Veronika Eyring\affil{1,2}}

\affiliation{1}{Deutsches Zentrum für Luft- und Raumfahrt (DLR), Institut für Physik der Atmosphäre, Oberpfaffenhofen, Germany}
\affiliation{2}{University of Bremen, Institute of Environmental Physics (IUP), Bremen, Germany}
\affiliation{3}{Computational Research Division, Lawrence Berkeley National Laboratory, Berkeley, CA, USA}
\affiliation{4}{Center for Climate Change and Policy Studies, Boğaziçi University, Istanbul, Turkey}
\correspondingauthor{Aytaç Paçal}{aytac.pacal@dlr.de}

\begin{keypoints}

\item Extreme temperature events are detected with Gaussian Mixture Models to follow a multimodal rather than a unimodal distribution.
\item 10-year temperature extremes will occur 13.6 times more frequently under 3.0$^\circ$C future warming.
\item Colder days are getting warmer faster than hotter days in high latitudes, whereas it is the opposite for many regions in low latitudes.
\end{keypoints}

\begin{abstract}
Extreme temperature events have traditionally been detected assuming a unimodal distribution of temperature data. We found that surface temperature data can be described more accurately with a multimodal rather than a unimodal distribution. Here, we applied Gaussian Mixture Models (GMM) to daily near-surface maximum air temperature data from the historical and future Coupled Model Intercomparison Project Phase 6 (CMIP6) simulations for 46 land regions defined by the Intergovernmental Panel on Climate Change (IPCC). Using the multimodal distribution, we found that temperature extremes, defined based on daily data in the warmest mode of the GMM distributions, are getting more frequent in all regions. Globally, a 10-year extreme temperature event relative to 1985-2014 conditions will occur 13.6 times more frequently in the future under 3.0$^\circ$C of Global Warming Levels (GWL). The frequency increase can be even higher in tropical regions, such that 10-year extreme temperature events will occur almost twice a week. Additionally, we analysed the change in future temperature distributions under different GWL and found that the hot temperatures are increasing faster than cold temperatures in low latitudes, while the cold temperatures are increasing faster than the hot temperatures in high latitudes. The smallest changes in temperature distribution can be found in tropical regions, where the annual temperature range is small. Our method captures the differences in geographical regions and shows that the frequency of extreme events will be even higher than reported in previous studies.
\end{abstract}

\section*{Plain Language Summary}
Extreme temperature events are unusual weather conditions with exceptionally low or high temperatures. Traditionally, the temperature range was determined by assuming a single distribution, which describes the frequency of temperatures at a given climate using their mean and variability. This single distribution was then used to detect extreme weather events. In this study, we found that temperature data from reanalyses and climate models can be more accurately described using a mixture of multiple Gaussian distributions. We used the information from this mixture of Gaussians to determine the cold and hot extremes of the distributions. We analysed their change in a future climate and found that hot temperature extremes are getting more frequent in all analyzed regions at a rate that is even higher than found in previous studies. For example, a global 10-year event will occur 13.6 times more frequently under 3.0$^\circ$C of global warming. Furthermore, our results show that the temperatures of hot days will increase faster than the temperature of cold days in equatorial regions, while the opposite will occur in polar regions. Extreme hot temperatures will be the new normal in highly populated regions such as the Mediterranean basin.

\section{Introduction} \label{sec:intro}
Increasing levels of atmospheric carbon dioxide (CO$_2$) concentration unequivocally transformed the earth's climate \cite{IPCC2021}. This surplus of CO$_2$ in the atmosphere contributes to the greenhouse effect, and by increasing the mean and the variability of global temperatures, it amplifies the risk of high-impact temperature extremes \cite{Baker2018}. The effects of anthropogenic global warming led to the emergence of heat extremes that would not have occurred previously \cite{Robinson2021}. This means that unprecedented heat extremes like the 2010 Russian heatwave or the 2021 Western North America heatwave would have likely not happened without the warming effect \cite{Rahmstrof2011, Christidis2015, Thompson2022}. The latter was found to be a remarkable four standard deviations away from the mean \cite{Thompson2022}. The Intergovernmental Panel on Climate Change (IPCC) Sixth Assessment Report (AR6) concluded that human influence on the climate system is unequivocal \cite{IPCC_2021_WGI_Ch_3} and \textit{virtually certain} to be the main driver of the changes in hot and cold extremes \cite{IPCC_2021_WGI_Ch_11}. It introduced more frequent and intense hot extremes since the 1950s on land areas while a decrease in cold extremes is observed \cite{IPCC2021}. Several studies found that the duration, frequency, and intensity of extreme events will increase, and extreme events will be introduced at new locations \cite{SREXch3, Rahmstrof2011, Kharin2013, Sillmann2013a, Sillmann2013b, Pfleiderer2019, PerkinsKirkpatrick2020, Vogel2020, Raymond2020, IPCC_2021_WGI_Ch_11, Mallick2022}. As the number of occurrences of heat extremes like the 2003 European heatwave and their duration increase, the socio-economic burden of climate change poses a threat to societies \cite{Meehl2004, Robine2008, Garcia2021, Demiroglu2020, Perera2020, IPCC_2021_WGI_Ch_11}.

The warming of the climate causes different changes in different regions. Tropics, polar regions and the Middle East and North Africa (MENA) region, are hot spots of notable climate trend shifts \cite{Hao2018, Zhang2022}. \citeA{Iyakaremye2022} have shown that an abrupt shift in the daily maximum temperatures occurred in Africa in the last two decades compared to the previous 20 years, which introduced more frequent and intense hot days. Moreover, regions in Africa will face a higher increase in temperatures compared to the rest of the globe. \citeA{Iyakaremye2021} found that the annual maximum of daily maximum temperatures over Africa is expected to increase by 1.6/2.2$^\circ$C in the future, while global temperatures are projected to rise by 1.5/2.0$^\circ$C during the same period. In the MENA region, the frequency and intensity of heatwaves will highly increase by the end of the century under a business-as-usual pathway scenario, which will affect about half of the MENA population \cite{Lelieveld2016, Zittis2021, Ozturk2021}. The number of occurrences of exceptionally hot summers, which have 2-4$^\circ$C hotter temperatures than the long-term average, has also increased from a single event between 1951 and 1980 to five events between 2001 and 2010 in Central and Eastern Europe, where the 2010 heatwave was the hottest and longest event with the largest geographical extent that ever occurred over Europe \cite{Twardosz2013, Guerreiro2018}. Similarly, other studies also found that the temperature extremes in Europe will increase 20-fold at the end of the century, compared to 1961-1990 \cite{Nikulin2011, Schar2004, Barriopedro2011}. Over the Americas, the dry and hot extremes showed an increase both in frequency and spatial scope over the past 122 years \cite{Alizadeh2020, Cai2014}. 

Correctly characterizing the temperature distributions to analyze extreme events is a still-continuing issue as extremes are by definition rare events, and several studies showed that the assumption of distributions or a stationary climate often underestimates the observed heat records \cite{Benestad2004, Schar2004, Anderson2010, Fischer2010, Barriopedro2011,  Li2019, Loikith2019}. \citeA{Thompson2022} characterized extreme events by calculating a daily extreme index which is the difference between the daily maximum temperature and mean daily maximum temperature divided by the standard deviation. With the assumption of a normal distribution, they found that the 2021 North American heatwave was one of the most extreme events with 4 standard deviations from the mean. Moreover, the authors projected that 20\% of the weather risk attribution forecast regions \cite{WRAF2019} will experience extreme events that are four standard deviations from the means in the future. Other studies found that hot summers will be the norm, i.e. mean temperatures exceed the temperature of the historically hottest summer, within the next 1-2 decades \cite{Mueller2016, Lewis2017, Vogel2020, Vogel2020a}. 

Common indices to monitor and analyze climate extremes that are used in the climate community at the moment, such as ETCCDI (the Expert Team on Climate Change Detection and Indices), are mostly based on daily mean near-surface air temperature or daily maximum near-surface air temperature \cite{Zhang2011, Alexander2006}. Two standard approaches to detect extreme events are the percentile-over-threshold (POT) and the block maxima method. The block maxima method groups data into an equal length of blocks, e.g. month, season, or year, and use the maximum temperature value of each block to fit the data. The POT method defines a threshold, e.g. percentiles, and uses all temperature values above this threshold in the analysis. Choosing the percentiles for defining extremes is not trivial as the temperature extremes have a strong seasonality and temporal dependence \cite{Huang2016}. The block maxima method is more commonly used in climate studies because of its simplicity with monthly, seasonal or annual block periods for fitting generalized extreme value (GEV) distribution to temperature and precipitation extremes \cite{Kharin2013, Wang2016, Paciorek2018, Wehner2018, Alaya2020, Li2021, IPCC2021}. The block maxima method, however, does not use all available data, as calculating a single maximum value from a block period throws out the rest of the data. To be approximated by the GEV distribution, the blocks are assumed to be long enough and ``max-stable", which means that if you take the maximum of a group of values selected from a specific GEV distribution, the result will be GEV distributed with the same shape parameter \cite{Huang2016, Alaya2020}. However, these assumptions might not be valid for all possible use cases or all possible variables. For example, GEV is not the best fit for shorter block lengths as the fit improves with increasing block size \cite{Alaya2020, Wang2016}. \citeA{Alaya2020} argued that the identically distributed random variables assumption of extreme value theory might be problematic for extreme precipitation events. They considered a mixture of GEV distributions to fit precipitation data to demonstrate that the mixture distribution could be a potential explanation for the instability of annual maxima. \citeA{Kollu2012} tested wind speed characteristics using mixture probability distribution functions (PDF). They found that conventional PDFs are inadequate to describe wind speed distributions compared to the mixture distributions that they used in the study. A mixture of Gaussians was used by \citeA{Shin2022} to describe the distribution of the daily thermal comfort index in South Korea, an index that has a strong seasonality. Ice surface temperature data follows a clear multimodal distribution, according to \citeA{Clarkson2022}. They also found that a unimodal distribution fit is particularly poor at modelling the tail probabilities. Probability distributions with one and two components are called unimodal and bimodal, respectively, whereas distributions with multiple (two or more) components are called multimodal distributions.

The temperature distributions are expected to move towards warmer temperatures and to change their shape with changing means and standard deviations \cite{IPCC2021}. Also, the assumption of distribution might not be correct for all geographical regions as daily weather variables show a distinct non-Gaussianity \cite{Volodin2012, Perron_Sura_2013, kodra_asymmetry_2014, Sardeshmukh2015, Linz2018, Tamarin2019}. Furthermore, several studies found that daily mean, daily maximum and real forecast data of 2m temperatures show bimodal features \cite{Grace1995, Wilks2002, Donat2012, Hongyeon2016, Bertossa2021}. These changes, shifts and bimodalities in the temperature distributions affect the probabilities in the tails. As extreme events are rare events that lie in the tails of a distribution, correctly describing the tails is very important for extreme event detection. Even though the block maxima method is widely used in studies which used block sizes large enough to converge asymptotically to GEV distributions, a GEV distribution is not well suited to describe extreme value data when the bimodality is apparent or block sizes are short \cite{Sardeshmukh2015, Wang2016, Knoben2019, Alaya2020}. Therefore, the properties of the entire probability distribution, i.e. mean, standard deviation and shape, are needed to get the tail properties right \cite{Sardeshmukh2015}. A distribution can be described by not only the mean and the standard deviation but also skewness and kurtosis. \citeA{Donat2012} found that daily minimum and maximum temperatures have significantly shifted towards higher values and skewed towards the hotter part of the distribution. They highlighted that the changes in extremes are related not only to the means but also to other parameters of the daily temperature distribution. \citeA{Sardeshmukh2009} found a parabolic relationship between kurtosis and skewness that cause the non-Gaussianity of the observed daily weather anomalies. Similarly, \citeA{Tamarin2022} used a mixture model with three Gaussians to describe the PDF of near-surface atmospheric temperature to analyze the relationship between kurtosis and skewness, as they are important to explain how the tails of the distribution change. They found that two- and three-Gaussian models are useful to explain the relationship between kurtosis and skewness. 

In the study presented here, our approach is to utilize the entire temperature distribution to detect extreme events. We implemented Gaussian Mixture Models (GMM), which describe the probability distribution function of data points as a mixture of Gaussian distributions. We determined the number of Gaussian components in the temperature distribution of each grid cell of 46 land regions defined by the Intergovernmental Panel on Climate Change (IPCC) using daily near-surface maximum air temperature data from the historical and future Coupled Model Intercomparison Project Phase 6 (CMIP6) simulations. This choice was supported by previous studies which found distinct bimodality in daily weather variables \cite{Grace1995, Wilks2002, Donat2012, Hongyeon2016, Bertossa2021} and was verified by applying the same analysis to the European Centre for Medium-Range Weather Forecasts Reanalysis 5th Generation (ECMWF-ERA5) data for the same historical time period (1985-2014). The parameters from the determined distribution components, namely mean, standard deviation and weight, were used to calculate the change in the return period of extreme temperature events between the historical and future periods determined by using global warming levels (GWL). In a stationary climate, the return period of an event describes the average time between the occurrences of a certain event of a defined size. In this study, we analysed 1-year, 5-year,  10-year and 20-year events, where an n-year event has an occurrence probability of 1/n as the climate is not stationary. Hence, these event magnitudes change as time progresses, where an \textit{n}-year event means that the event in question would be expected to occur once in every \textit{n} years.  We only calculated return periods equal to or less than the available future data period to prevent overestimating the return periods of extreme events, since GMM distributions are not bounded. Section \ref{sec:data} presents the climate data and warming levels used in this study, as well as the analyzed regions, and explains the methodology of detecting extreme event return periods by using GMM. Section \ref{sec:results} shows our results obtained using the GMM method for all analyzed IPCC land regions, and section \ref{sec:discussion} finalizes the paper with a summary and discussion.

\section{Data and Methodology} \label{sec:data}

\subsection{Climate Data}
\begin{figure}[!ht]
    \centering
    \includegraphics[width=0.8\textwidth]{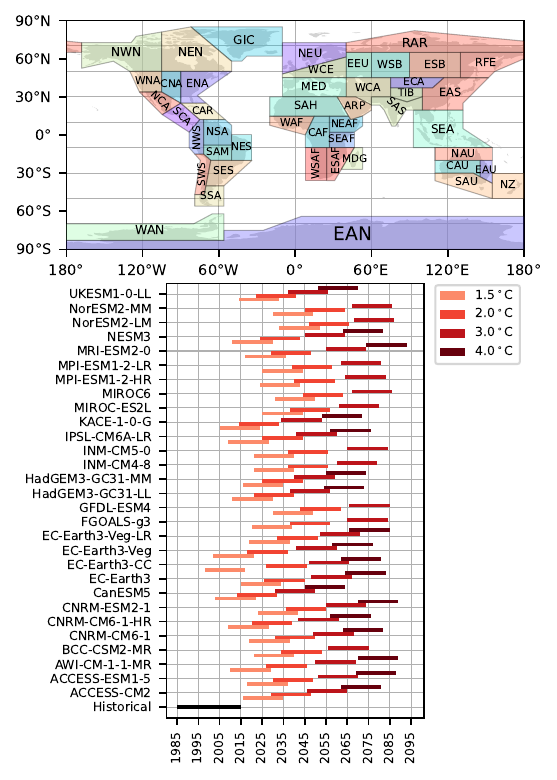}
    \caption{(Top) We used 46 land regions defined in \citeA{Iturbide2020}. See Supplementary Material Table S2 for region definitions. (Bottom) Future periods of the CMIP6 models when the central year of the 20-year running window exceeds global warming levels relative to 1850-1900 base for the SSP5-8.5 scenario are extracted using the data from \citeA{gwl_periods}. The colours in the graph go from light to dark, each colour representing a different level of warming 1.5$^\circ$C, 2$^\circ$C, 3$^\circ$C, and 4$^\circ$C. These levels are expected to be exceeded around 2026, 2040, 2060, and 2070 respectively. The 30-year historical base period is indicated in grey. Note that different models have different time periods when they exceed the GWL. Future periods for other SSP scenarios are presented in the Supplementary Material Table S3.}
    \label{fig:ipcc_regions_data_periods_585}
\end{figure}

For this study, we used multi-year daily near-surface maximum temperatures from the Coupled Model Intercomparison Project Phase 6 (CMIP6), and for which both the historical simulations and the simulations for Shared Socioeconomic Pathways (SSPs) 1-2.6, 2-4.5, 3-7.0 and 5-8.5 scenarios were available \cite{ONeill2014, Eyring2016, Oneill2016}. Additionally, the ECMWF-ERA5 dataset was included for the 30-year time period (1985-2014) \cite{ERA5_data}. Table \ref{tab:gcm} shows the list of models and their resolutions. The 30-year time period from 1985 to 2014 from historical simulations is used as the base to calculate the return values of extreme temperature events, i.e. 1-year, 5-year, 10-year and 20-year events. The GWL, as introduced in the IPCC AR6 report, are used to assess the changes in future climate in line with the warming levels defined in the Paris Agreement which are compared to the pre-industrial period \cite{IPCC2021}. The future period for each model is defined as a 20-year period between 2015 and 2100 when the central year of the running window of the global daily near-surface temperature mean of that model first exceeds 1.5$^\circ$C, 2$^\circ$C, 3$^\circ$C, and 4$^\circ$C relative to 1850-1900 global daily near-surface mean temperatures. We used the same GWL periods defined for and used in IPCC \cite{IPCC2021, gwl_periods}, similarly to \citeA{hajat2022} and \citeA{Ribeiro2022}. Therefore, we obtained the start and end years of 20-year GWL periods for each CMIP6 simulation from \citeA{gwl_periods}. Here, we used a longer historical base period (30 years) compared to future GWL periods (20 years) for the analysis to obtain more robust results. This decision was made based on the fact that GMM distributions have no bounds. Therefore, we focused our analysis solely on return periods shorter than our base period. By limiting our analysis to shorter return periods, we can mitigate the biases and outliers that may occur beyond the limits of the datasets. As some datasets did not exceed certain warming levels, they were excluded from the analysis (e.g NOR-ESM2-MM was not used in calculations for 4$^\circ$C warming under SPP5-8.5, as it did not exceed this level). Figure \ref{fig:ipcc_regions_data_periods_585} shows the historical and future GWL periods for each CMIP6 model used in this study.

\begin{table}[!ht]
\centering
\caption{Reanalysis data and CMIP6 models used in this study to detect extreme temperature events. Climate models with spatial resolutions ranging from 50 to 500 km were used in the analyses. The first available ensemble members were chosen. The Reanalysis dataset that has a resolution of 25km was regridded to 100 km and used for evaluating modality.}
\label{tab:gcm}
\begin{tabular}{lllll}
\toprule
Model            & Variant  & Resolution & Reference              &  \\ \midrule
ECMWF-ERA5       & Reanalysis  & 25 km      & \cite{ERA5_data}            &  \\
ACCESS-CM2       & r1i1p1f1 & 250 km     & \cite{ACCESS-CM2}      &  \\
ACCESS-ESM1-5    & r1i1p1f1 & 250 km     & \cite{ACCESS-ESM1-5}   &  \\
AWI-CM-1-1-MR    & r1i1p1f1 & 100 km     & \cite{AWI-CM-1-1-MR}   &  \\
BCC-CSM2-MR      & r1i1p1f1 & 100 km     & \cite{BCC-CSM2-MR}     &  \\
CanESM5          & r1i1p1f1 & 500 km     & \cite{CanESM5}         &  \\
CNRM-CM6-1       & r1i1p1f2 & 250 km     & \cite{CNRM-CM6-1}      &  \\
CNRM-CM6-1-HR    & r1i1p1f2 & 50 km      & \cite{CNRM-CM6-1-HR}   &  \\
CNRM-ESM2-1      & r1i1p1f2 & 250 km     & \cite{CNRM-ESM2-1}     &  \\
EC-Earth3        & r1i1p1f1 & 100 km     & \cite{EC-Earth3}       &  \\
EC-Earth3-CC     & r1i1p1f1 & 100 km     & \cite{EC-Earth3-CC}    &  \\
EC-Earth3-Veg    & r1i1p1f1 & 100 km     & \cite{EC-Earth3-Veg}   &  \\
EC-Earth3-Veg-LR & r1i1p1f1 & 250 km     & \cite{EC-Earth3-Veg-LR}&  \\
FGOALS-g3        & r1i1p1f1 & 250 km     & \cite{FGOALS-g3}       &  \\
GFDL-ESM4        & r1i1p1f1 & 100 km     & \cite{GFDL-ESM4}       &  \\
HadGEM3-GC31-LL  & r1i1p1f3 & 250 km     & \cite{HadGEM3-GC31-LL} &  \\
HadGEM3-GC31-MM  & r1i1p1f3 & 100 km     & \cite{HadGEM3-GC31-MM} &  \\
INM-CM4-8        & r1i1p1f1 & 100 km     & \cite{INM-CM4-8}       &  \\
INM-CM5-0        & r1i1p1f1 & 100 km     & \cite{INM-CM5-0}       &  \\
IPSL-CM6A-LR     & r1i1p1f1 & 250 km     & \cite{IPSL-CM6A-LR}    &  \\
KACE-1-0-G       & r1i1p1f1 & 250 km     & \cite{KACE-1-0-G}      &  \\
MIROC6           & r1i1p1f1 & 250 km     & \cite{MIROC6}          &  \\
MIROC-ES2L       & r1i1p1f2 & 500 km     & \cite{MIROC-ES2L}      &  \\
MPI-ESM1-2-HR    & r1i1p1f1 & 100 km     & \cite{MPI-ESM1-2-HR}   &  \\
MPI-ESM1-2-LR    & r1i1p1f1 & 250 km     & \cite{MPI-ESM1-2-LR}   &  \\
MRI-ESM2-0       & r1i1p1f1 & 100 km     & \cite{MRI-ESM2-0}      &  \\
NESM3            & r1i1p1f1 & 250 km     & \cite{NESM3}           &  \\
NorESM2-LM       & r1i1p1f1 & 250 km     & \cite{NorESM2-LM}      &  \\
NorESM2-MM       & r1i1p1f1 & 100 km     & \cite{NorESM2-MM}      &  \\
UKESM1-0-LL      & r1i1p1f2 & 250 km     & \cite{UKESM1-0-LL}     &  \\ \bottomrule
\end{tabular}
\end{table}

We extracted daily maximum near-surface air temperature for 30-year historical and 20-year future periods under GWL for each SSP individually for 46 IPCC land regions that are shown in Figure \ref{fig:ipcc_regions_data_periods_585} \cite{Iturbide2020}. All data extraction and preprocessing in this study were performed by using the Earth System Model Evaluation Tool (ESMValTool) version 2.5.0, which is an open-source software package for analysing and evaluating model simulations \cite{Eyring2020, Lauer2020, Righi2020, Weigel2020}. We extracted the daily maximum near-surface air temperature from each model for each region using shapefiles provided by IPCC \cite{Iturbide2020}, converted units from Kelvin to Celsius, and created a single spatiotemporal Network Common Data Form (NetCDF) file for each region. The data were then ready to be used in the diagnostic script written in Python where the extreme events and their return periods were analyzed. 

\subsection{Return Period Analyses}  \label{sec:GMM}

For the return period calculation of extreme temperature events, i.e. 1-year, 5-year, 10-year and 20-year events, we defined a temperature threshold for an event by calculating the standard deviation distance of the event temperature from the mean temperature in the past, i.e. how many standard deviations away the event temperature \textit{was} from the mean. We then applied this temperature threshold value to the future period but calculated its standard deviation distance from the mean using the parameters from the future distribution, i.e. how many standard deviations away the event temperature \textit{will be} from the mean. To test the underlying distribution shape of the daily near-surface maximum temperature distribution, we first analyzed data from individual grid cells of each climate model. We found that daily maximum near-surface air temperature data in climate grid cells usually do not follow a unimodal distribution, but rather follow a bimodal distribution, a probability distribution composed of two components.

\begin{figure}[!ht]
        \centering
        \includegraphics[width=0.85\textwidth]{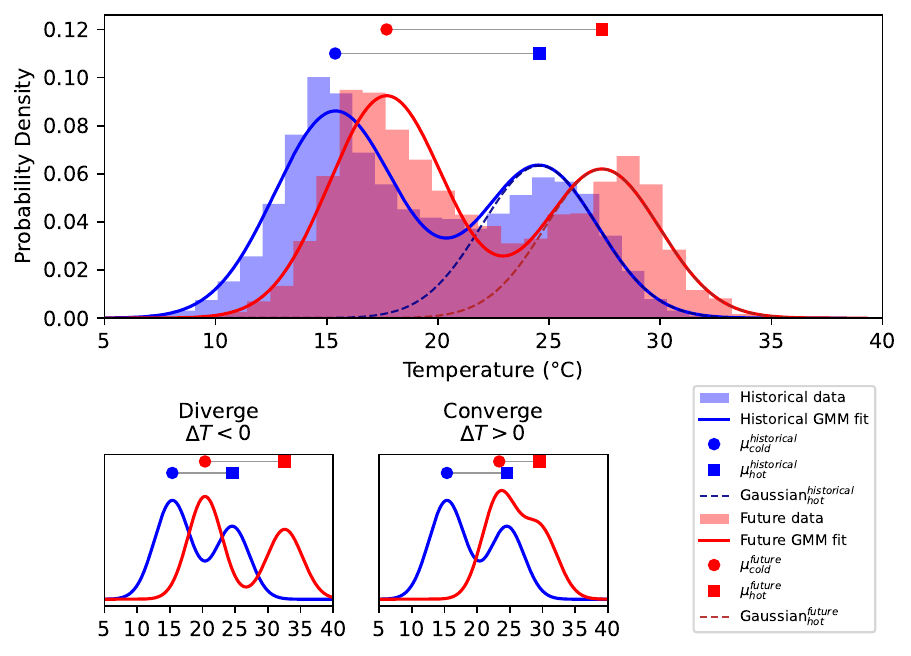}
        \caption{Exemplary bimodal distribution of daily maximum temperatures from a grid cell for the historical 30-year period of 1985-2014 (blue) and future 20-year GWL period (red). Blue and red lines show the corresponding GMM fit for the historical and future periods, respectively. The shape of the distribution is determined by the parameters of each Gaussian component, which are the means, standard deviations and weights. Here, the means of cold and hot Gaussian peaks are shown with blue(red) dots and squares for the historical (future) period, respectively. The hot Gaussian component used in the analysis is shown with a dashed blue (red) line for the historical (future) period. The bottom two plots show what convergence and divergence of the peaks mean based on the $\Delta T$ value.}
        \label{fig:gmmexplained}
\end{figure}

To calculate the return periods of extreme events, we modelled the probability distribution of multi-year (30 years for the historical base period and 20 years for future GWL periods) daily near-surface maximum temperature data from a grid cell as mixtures of multiple Gaussian distributions, rather than a single Gaussian distribution. GMM is a probabilistic model that describes the data points in a population as a mixture of Gaussian distributions with unknown parameters which are the mean, standard deviation and weight of each Gaussian component, five parameters in total for a bimodal distribution. With this approach, we were able to analyse the change in the distribution, and accurately model the tails of the data compared to a unimodal distribution. When a unimodal distribution is fit to multi-year data with bimodality, it is likely that the resulting distribution will have a larger standard deviation to encompass both modes. This large standard deviation between a unimodal distribution fit to bimodal data can have significant implications for analyses, as the larger standard deviations of unimodal distributions tend to push the extreme events further away from what would be observed if the bimodality were properly accounted for. In other words, when one tries to calculate the threshold of an event as \textit{n} sigma distance from the mean, this threshold might be well beyond the maximum value of the distribution. Furthermore, a unimodal distribution fit will affect the measures of central tendency when bimodality exists in data \cite{gmm_mean_median}. An example goodness-of-fit test for normal distribution, GEV distribution with different shape parameters and GMM distributions on the daily maximum temperature data from a random grid cell is presented in Supplementary Material Section 1. We used an unsupervised machine-learning package, the ``GaussianMixture" package from open-sourced machine-learning library Scikit-learn, to compute the unknown parameters of the Gaussian components in a mixture that generates all observed data points \cite{scikit-learn}. We applied this package to the daily maximum near-surface air temperature data in each grid cell of the CMIP6 models. The ``GaussianMixture" package first randomly assigns values to the component parameters and then uses the expectation-maximisation algorithm (EM) to converge their values. EM algorithm fits GMM to data by alternating between two steps, Expectation (E) and Maximisation (M). In the E step, it randomly assumes components and calculates the probability of each point to be generated by that component. In the M step, the parameters are tweaked to maximise the likelihood found in the first step. It also uses the Bayesian Information Criteria (BIC) score, which is used to estimate the goodness-of-fit of a distribution and which accounts for both the likelihood function and the number of parameters. Then, the probability distribution function of the mixture model that was fit to multi-year daily near-surface temperature can be written as a linear summation of multiple Gaussian components:

\begin{linenomath*}
\begin{align}
    p(x) &= \sum_{k=1}^K \omega_k \mathcal{N}( x \mid \mu_k, \sigma_k ) \\
    \mathcal{N}(x \mid \mu_k, \sigma_k ) &= \frac{1}{\sigma_k{\sqrt{2\pi}}} \exp{ \left(-\frac{(x-\mu_k)^2}{2\sigma_k^2}\right)} \\
    \sum_{k=1}^K \displaystyle \omega_k &= 1 
\end{align}
\end{linenomath*}

where $K$ is the number of Gaussian components in the mixture. $\mu_k$, $\sigma_k$ and $\omega_k$ are the mean, the standard deviation and the weight of the $k^{th}$ component, respectively. Implementing Gaussian mixture models to evaluate multi-year raw daily maximum temperatures allows us to investigate the long-term characteristics of the individual components. This method does not consider the temporal changes within one period, as they can be assumed to be negligible compared to the changes between different time periods. As shown in other studies, mean temperatures are increasing all over the globe \cite{IPCC2021, Robinson2021, Eyring2020}. Using the raw temperatures, we can analyse how the convergence or divergence of the peaks of the different Gaussian components affect the extremes compared to the used historical periods. In our analysis, we have disregarded three or more Gaussian components. This choice was supported by the value of the BIC score and the fact that increasing the number of components tends to cause overfitting, even though BIC scores penalise adding more parameters. In some cases, the BIC scores for the components showed close results for more than three components (see Figure S2 in the Supplementary Material). For instance, the lowest BIC score was reached for a mixture with seven Gaussian components for the distribution of temperatures in a grid cell. However, the highest change in BIC scores occurred when switching from one component to two components. Consequently, we used the gradient of BIC scores rather than using the lowest score. We selected the number of Gaussian components where the highest gradient change occurs in the BIC scores as the best fit. To further prevent overfitting, we also applied the following unimodality test after estimating the BIC scores: If the BIC score returned a bimodal distribution, then the parameters of the mixture distribution components were used for the unimodality test. As shown in Equation \ref{eqn:unimodality}, if the difference between the means of Gaussian components was less than or equal to twice the minimum of standard deviations, then unimodal distribution was assumed, otherwise, the bimodal distribution fit for the data was kept. It is worth noting that this procedure had a tendency to favour fitting a unimodal distribution. However, after all these tests and checks, the majority of grid cells showed a clear bimodal distribution. For a bimodal distribution, hereafter we referred to the right (left) Gaussian component as ``hot (cold) Gaussian" as shown in Figure \ref{fig:gmmexplained}).
\begin{linenomath*}
\begin{equation}
    |\mu_1-\mu_2| 	\leq 2 \min (\sigma_1, \sigma_2)
\label{eqn:unimodality}
\end{equation}
\end{linenomath*}

\begin{figure}[!ht]
    \centering
    \includegraphics[width=0.95\textwidth]{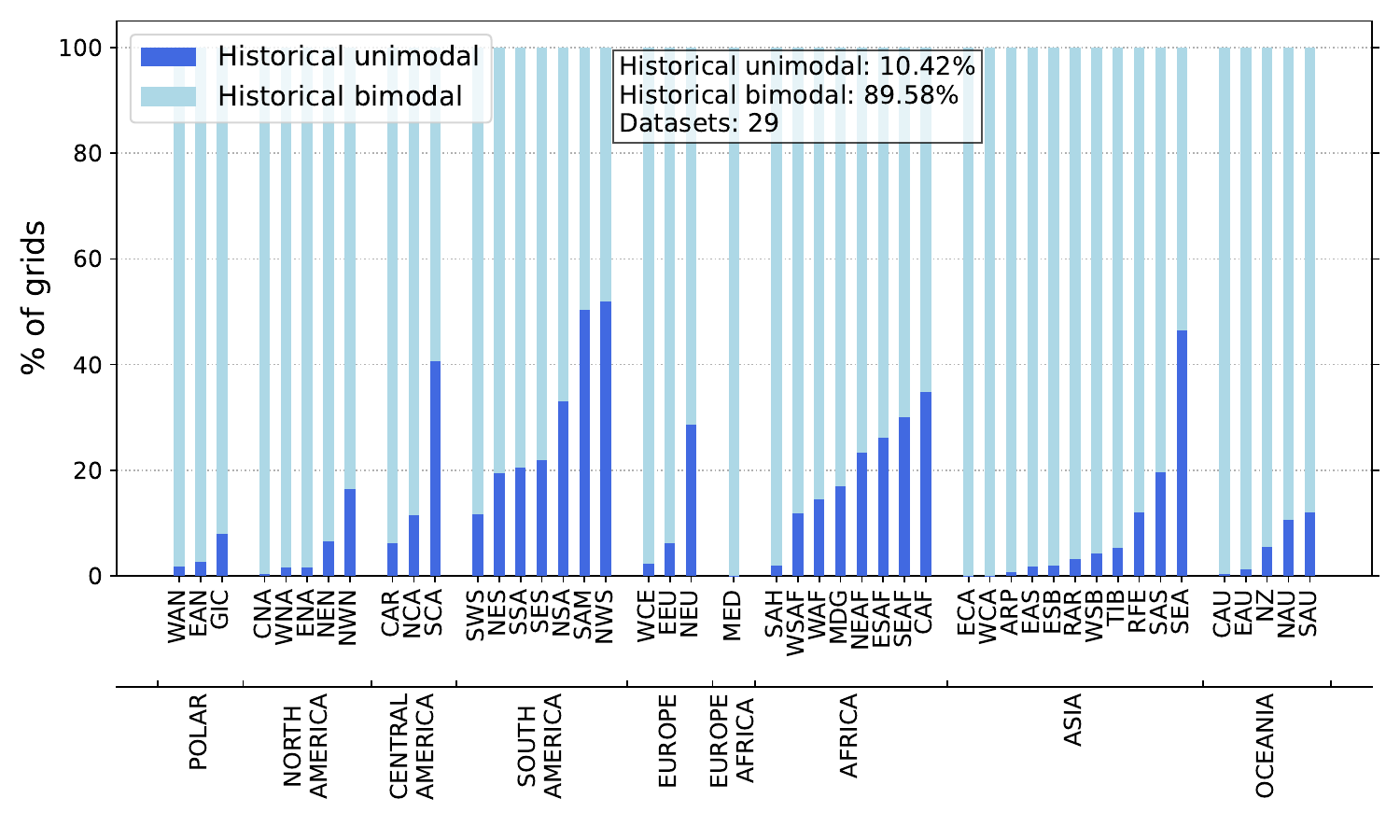}
    \caption{Multi-model mean percentages of grid modalities for the historical period in study regions grouped by continents. Dark and light blue bars show the percentage of grid cells with unimodal or bimodal distribution, respectively, for the historical period of 29 CMIP6 simulations.}
    \label{fig:modality_hist}
\end{figure}

\begin{figure}[!ht]
    \centering
    \includegraphics[width=0.95\textwidth]{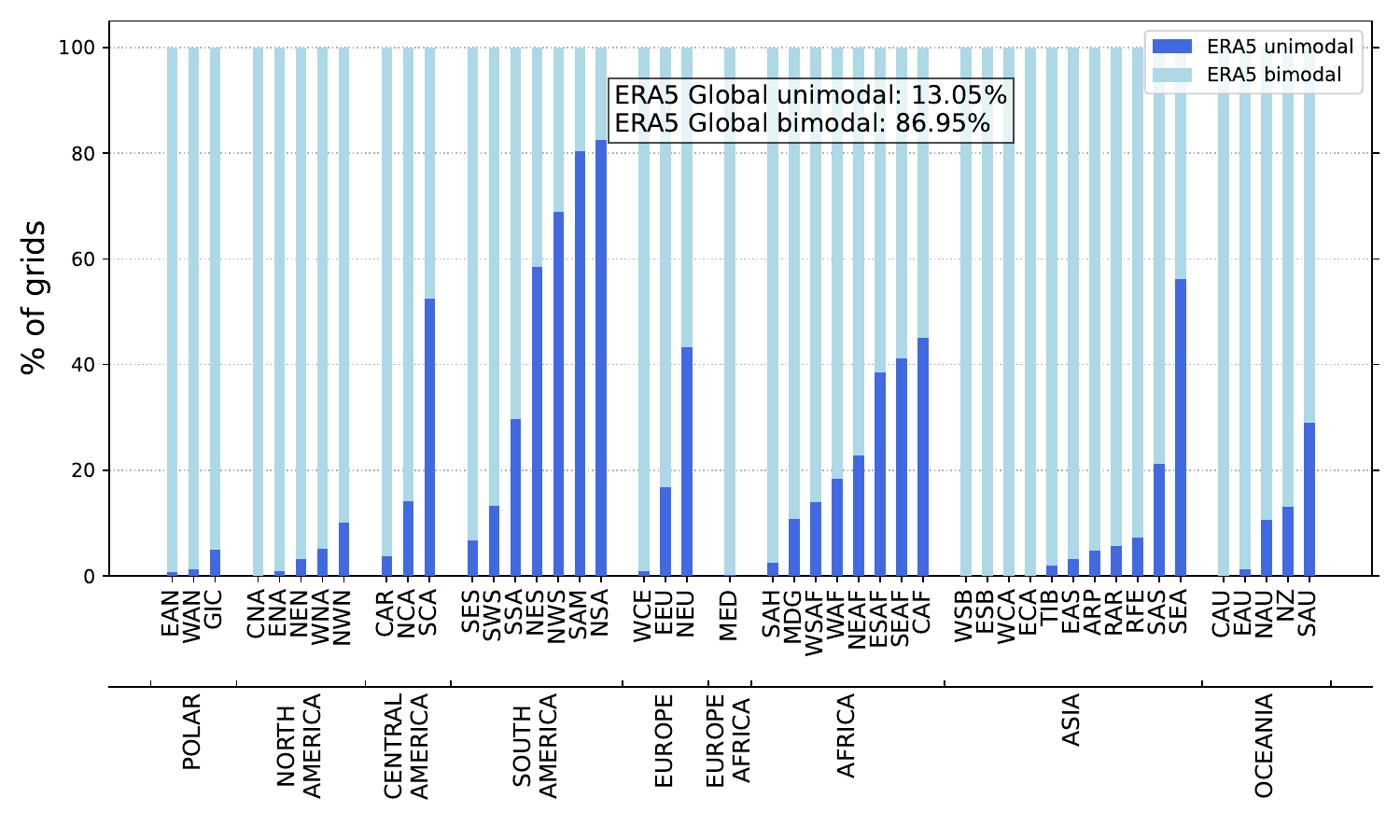}
    \caption{Same as Figure \ref{fig:modality_hist} but for ECMWF-ERA5 reanalysis dataset.}
    \label{fig:modality_era5}
\end{figure}

First, we grouped grid cells of a region depending on their modality, either unimodal or bimodal, for each CMIP6 model, and calculated the percentages of grid modalities among all grid cells of a region for each CMIP6 model. We then determined the multi-model mean percentages of grid cell modalities of a region as shown in Figure \ref{fig:modality_hist}. Additionally, we calculated the global multi-model mean percentage of grid cell modalities using all regions and CMIP6 models. We found that globally 88.78\% of all grid cells follow a bimodal distribution in the historical period as shown in the white box in the upper centre part of Figure \ref{fig:modality_hist}. Furthermore, we analysed the ECMWF-ERA5 dataset for the same historical time period (1985-2014) to confirm whether bimodality is also found in data other than model simulations. We regridded the ECMWF-ERA5 data from a 25-km grid to a coarser 1-degree 100-km grid using the nearest neighbour method to have a similar resolution as many CMIP6 datasets. The ECMWF-ERA5 reanalysis dataset shows similar results to the CMIP6 models: Globally 86.95\% of all grid cells in the ECMWF-ERA5 reanalysis dataset follow a bimodal distribution as shown in the white box in the upper centre part of Figure \ref{fig:modality_era5}, while only 13.05\% of them follow a unimodal distribution. The temperature distributions in the ERA5 and CMIP6 datasets show predominantly similar patterns across various regions, although certain exceptions are observed, particularly in South America. These differences can most likely be attributed to several factors. First, resampling of ERA5 data from a 25 km grid to a coarser 1-degree grid introduces a smoothing effect on the data, which would increase the unimodal grid cells. Additionally, biases in surface temperature in CMIP6 datasets also contribute to the observed variations from ERA5 \cite{Bock2020}. Nevertheless, as we aim to evaluate the shape of temperature distributions, we did not apply a bias correction and used raw multi-year daily temperature data from CMIP6 models for our analysis.


Then, the parameters of the hot Gaussian component, $\mu_{hot}^{historical}$, $\sigma_{hot}^{historical}$ and $\omega_{hot}^{historical}$, were used to calculate the change in return periods. We only analysed 1-year, 5-year, 10-year and 20-year events, as GMM are unbounded. One should be careful while calculating the return periods using GMM, as the unbounded tails of the Gaussian component could overestimate the probabilities of longer return periods. Therefore, return periods equal to or less than the analysis period were calculated using GMM. The change in return periods is calculated first in each grid cell of a region and then averaged together to produce regional results for each CMIP6 simulation.

For normally distributed data, the expected percentage of the population inside the $\mu\pm d\sigma$ range is defined as
\begin{linenomath*}
\begin{equation}
    E(\mu\pm d\sigma) = \erf\bigg(\frac{d}{\sqrt{2}}\bigg)
\end{equation}
\end{linenomath*}
where $\erf$ is the error function and $d$ is the standard deviation distance. The approximate expected frequency, $f$, outside this range is then defined as the return period of an extreme.
\begin{linenomath*}
\begin{equation}
    1\ in\ \frac{1}{1-\displaystyle \erf\bigg(\frac{d}{\sqrt{2}}\bigg)}\ days
\label{eqn:expectedfreq}
\end{equation}
\end{linenomath*}

The return period of an event describes the average time between the occurrences of a certain event of a defined size, where an $n-year$ event has an occurrence probability of 1/n as the climate is not stationary, where ``\textit{year}" is defined as the number of days covered by the hot Gaussian component. The reason for this definition is that the entire probability distribution is composed of both the cold and warm periods of a year, however, our dataset consists of daily maximum temperature data spanning 30 (or 20) years, totalling 10,950 (or 7,300) days. Since our analysis specifically aims to identify extreme values using parameters from the hot Gaussian component, we need to consider the number of data points generated by this component as the definition of a "year". We determine the length of a "year" by dividing the data points falling under the hot Gaussian component by the length of the analysis period as shown in Equation \ref{eqn:yeardef}. For example, we can assume that a symmetrical bimodal distribution results in $\sim$180 days of cold weather and $\sim$180 days of hot weather in a normal 365-day calendar year. For such a symmetric case, a 10-year event would then be a temperature event in 1800 days (10 years$\textstyle\times 180 \textstyle \frac{days}{year}$). Since we cannot assume a symmetric distribution for grid cells of each model, we calculated the number of days covered by the hot Gaussian component using the component weights and dataset size.

Let $\mathcal{D}$ denote the number of days in $L$ years. Then, a ``\textit{year}" in the historical period, $\mid \mathcal{N}(\mu_{hot}^{historical}, \sigma_{hot}^{historical}) \mid$ is defined as
\begin{linenomath*} 
\begin{equation}
\label{eqn:yeardef}
    \mid \mathcal{N}(\mu_{hot}^{historical}, \sigma_{hot}^{historical} )\mid = 
    \frac{\omega_{hot}^{historical} \mathcal{D}} {L}
\end{equation} 
\end{linenomath*}
where $\mu_{hot}^{historical}$ is the mean, $\sigma_{hot}^{historical}$ is the standard deviation and $\omega_{hot}^{historical}$ is the weight of hot Gaussian component. The expected frequency of \textit{n}-year events in the historical period, $f^{historical}_{n}$, is then calculated by using the length of a year,
\begin{linenomath*} 
\begin{equation}
    f^{historical}_{n} = n \times \mid \mathcal{N}(\mu_{hot}^{historical}, \sigma_{hot}^{historical}) \mid \quad  n = 1, 5, 10, 20
\label{eqn:freqhist}
\end{equation} 
\end{linenomath*}
The standard deviation distance of range, $d^{historical}_n$, for an extreme event in the historical period can be calculated by using Equation \ref{eqn:expectedfreq},
\begin{linenomath*} 
\begin{equation}
    d^{historical}_n = \displaystyle \erf^{-1}\bigg(1-\frac{1}{f^{historical}_n}\bigg)\sqrt{2}
\label{eqn:sigmarange}
\end{equation} 
\end{linenomath*}
where $\erf^{-1}$ is inverse error function. Now, we can calculate a temperature threshold, $\tau^{historical}_n$, for an \textit{n-year} event in the historical period.
\begin{linenomath*} 
\begin{equation}
\tau^{historical}_n = \mu_{hot}^{historical}+d^{historical}_n\sigma_{hot}^{historical}
\label{eqn:templimit} 
\end{equation} 
\end{linenomath*}
Using this temperature threshold from the historical period, we calculate the standard deviation distance of the temperature threshold of \textit{n-year} event in the future, $d^{future}_n$, by using the mean $\mu^{future}_{hot}$, and standard deviation $\sigma^{future}_{hot}$ from the hot Gaussian component of the future distribution.
\begin{linenomath*} 
\begin{align}
d^{future}_n &= \frac{\tau^{historical}_n-\mu^{future}_{hot}}{\sigma^{future}_{hot}} \\
f_{\dot{n}}^{future} &= \frac{1}{1-\erf\bigg(\displaystyle \frac{d^{future}_n}{\sqrt{2}}\bigg)} \\
\end{align}
\end{linenomath*}
Finally, the new value of the return period in the future $\dot{n}$, i.e. $\dot{n}$-year event, is calculated by using Equation \ref{eqn:freqhist}
\begin{linenomath*}
\begin{equation}
\label{eqn:nyear}
   \dot{n} = \frac{f_{\dot{n}}^{future}}{\vert \mathcal{N}(\mu_{hot}^{future},\sigma_{hot}^{future}) \vert}
\end{equation}
\end{linenomath*}
where $\mathcal{N}(\mu_{hot}^{future}, \sigma_{hot}^{future}) \vert$ is length of a ``\textit{year}" in the future period.

With this method, we can also analyse if and how much the Gaussian components will shift in the future relative to the historical period. We defined $\Delta T$, as the change in the difference between the means of cold and hot Gaussian components as shown in Equation \ref{eq:deltat}:
\begin{linenomath*} 
\begin{align}
\label{eq:deltat}
\Delta T &= \delta T_{cold} - \delta T_{hot} \\
\delta T_{cold} &= \mu_{cold}^{future}-\mu_{cold}^{historical} \\
\delta T_{hot} &= \mu_{hot}^{future}-\mu_{hot}^{historical}
\end{align} 
\end{linenomath*}
In Figure \ref{fig:gmmexplained}, this change in hot and cold Gaussian means is schematically illustrated. Assuming the future means of Gaussian components are higher than the historical periods, $\delta T_{cold}$ and  $\delta T_{hot}$ will always be positive. Therefore, a negative $\Delta T$ means that the peaks are diverging in the future: the hot Gaussian moves toward warmer temperatures faster than the cold Gaussian, which increases the frequency of hot extremes and induces an overall warmer climate. A positive $\Delta T$ means that the peaks are converging: the cold Gaussian moves closer to the hot Gaussian, which increases the number of days with warmer temperatures in the colder mode.

\section{Results} \label{sec:results}

First, we checked the change in the percentage of modalities from the present to the future time periods. For this, we analyzed the modality of the temperature data from each individual grid cell of an IPCC land region by counting the number of grid cells with each modality. We found that the percentages of grid cells with bimodal distributions stay almost the same under different warming levels. As some of the CMIP6 datasets do not exceed certain warming levels, the number of datasets are not identical for the historical and future period and therefore affect the change in percentages. We analysed modalities of grid cells under different GWL for all SSP scenarios but we only present SSP5-8.5 results here, as the SSP5-8.5 scenario had data from 29 CMIP6 models and the GWL is scenario independent. Globally, almost 90\% of all grid cells follow a bimodal distribution as shown in Figure \ref{fig:modality_hist} for the historical period, Figure \ref{fig:modality_era5} for the reanalysis data and Figure \ref{fig:modality_58530} for GWL 3.0$^\circ$C for different regions grouped by continents (See Supplementary Material Table S3 for other warming levels). Global averages and the number of datasets are shown in the white box in the upper centre part of each figure. In the historical period, the grid cells in tropical and sub-tropical regions have slightly higher percentages of unimodal distributions compared to higher latitude regions. However, regions still mostly follow a bimodal distribution as shown in Figure \ref{fig:modality_hist}. The multi-model mean percentage of unimodal distributions does not exceed 50\% of grid cells in any of the regions, except in N.W.South-America (NWS) and South-American-Monsoon (SAM) regions where 51.94\% and 50.33\% of the grid cells follow a unimodal distribution, respectively, in the historical period. The higher percentage of unimodal distributions in lower latitudes is consistent with tropical climate features, where hot temperatures are observed all year round and the annual temperature range is small \cite{Richter2016, Beck2018}. This climate type is therefore expected to likely experience a temperature distribution close to a single Gaussian. All grid cells (99.9\%) in CMIP6 models follow a bimodal distribution in the Mediterranean (MED) region in the historical period and under all future periods. In polar regions, more than 90\% of the grid cells follow a bimodal distribution in the historical period. The percentage of grid cells with unimodal distributions in polar regions slightly increases under future global warming levels.

\begin{figure}[!ht]
    \centering
    \includegraphics[width=0.95\textwidth]{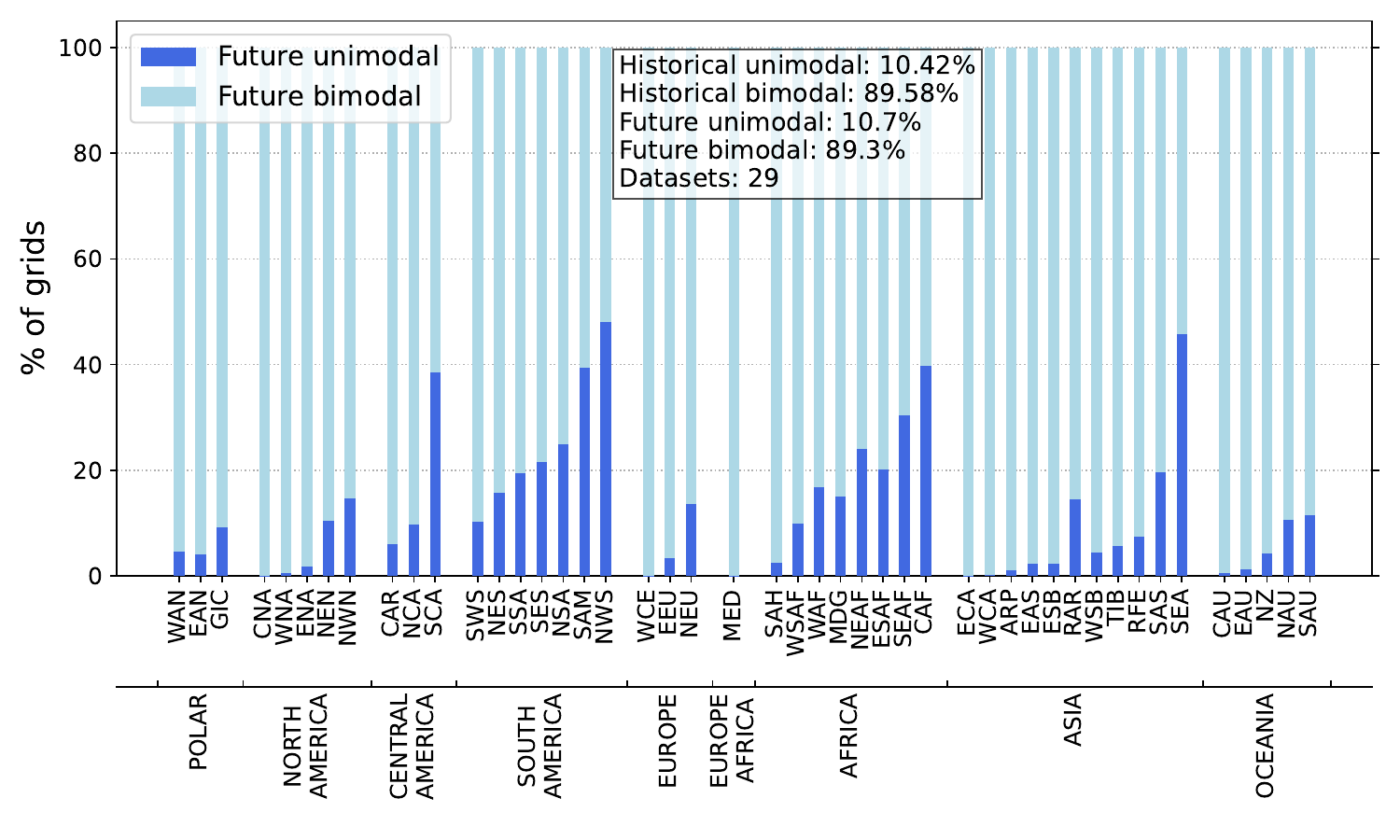}
    \caption{Same as Figure \ref{fig:modality_hist} but for future SSP5-8.5 scenario under GWL 3.0$^\circ$C.}
    \label{fig:modality_58530}
\end{figure}

\begin{figure}[!ht]
    \centering
    \includegraphics[width=\textwidth]{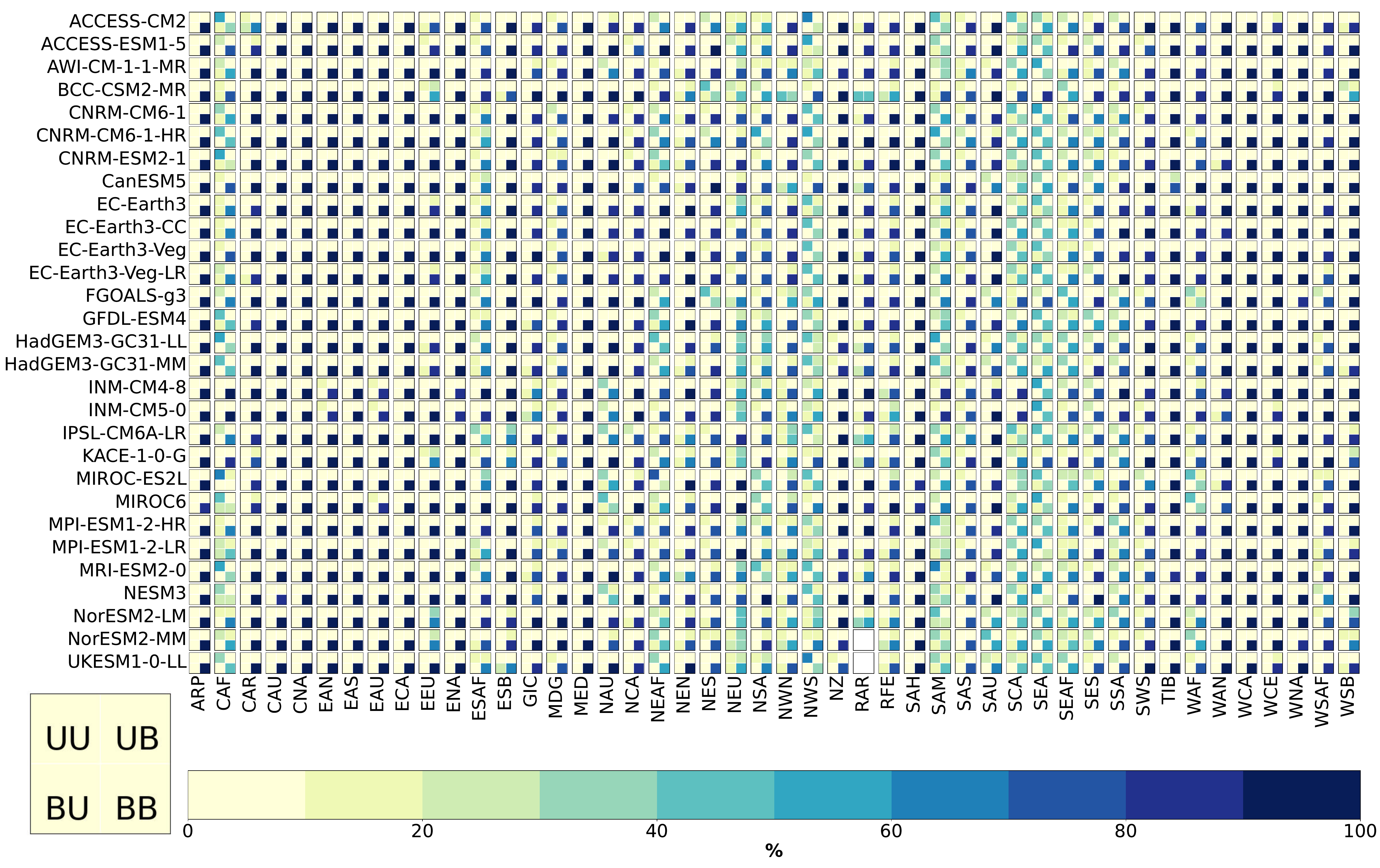}
    \caption{Percentage of changes in grid cell modalities relative to 1985-2014 distribution shape for SSP5-8.5 under GWL3.0$^\circ$C. Each cell represents a region of a CMIP6 model and is divided into 4 quadrants. Each quadrant of squares, $q_{ij}$, uses index notation, where $i$ represents the modality in the historical period and $j$ represents the modality in the future period, 1 for a unimodal distribution and 2 for a bimodal distribution. The top-left quadrant, $q_{11}$, shows the percentage of grid cells with unimodal distribution both in the historical and the future periods, i.e. unimodal to unimodal (UU). The top-right quadrant, $q_{12}$, shows the percentage of grid cells that change from unimodal distribution in the historical period to bimodal distribution in the future (UB). The bottom-left quadrant, $q_{21}$, shows bimodal to unimodal (BU). The bottom-right quadrant, $q_{22}$, shows bimodal to bimodal distribution (BB). The colour of the quadrants shows the percentage of grid cells. For several models, East Antarctica (EAN) region is not included in the analysis because it is composed of many grid cells near the pole, causing numerical problems.}
    \label{fig:gleckler_58530}
\end{figure}

As previously mentioned in Section \ref{sec:GMM}, large values of $\Delta T$ (see Equation \ref{eq:deltat}) will cause the temperature distribution to change its modality for future GWL periods with respect to the historical base period of 1985-2014. We analysed all regional grids for all CMIP6 models for the modality changes under GWL 1.5$^\circ$C, 2$^\circ$C, 3$^\circ$C, and 4$^\circ$C. Figure \ref{fig:gleckler_58530} shows the percentage of changes in grid cell distribution modalities under GWL3.0$^\circ$C. Globally, the percentage of grids changing from a unimodal (bimodal) distribution in the historical period to a bimodal (unimodal) distribution in the future periods is between 2.79\% (2.26\%) and 6.02\% (3.88\%) for different scenarios and GWL as shown in Table \ref{tab:modality_change}. The change from unimodal to bimodal distribution in the future period is most prevalent in regions where the highest percentage of unimodality was observed in the historical period, as shown in Figure  \ref{fig:modality_hist}. This suggests that regions that were previously characterized by more consistent temperatures (as indicated by a unimodal temperature distribution) may experience more variability in temperature in the future. As our analysis uses the mean and standard deviation of the same component from the historical and future daily maximum temperature distributions, we only used the grid cells which have the same modality in the historical and future periods. We disregarded the grid cells with changing modalities, i.e. unimodal to bimodal or vice versa, as this will affect the mean and standard deviation, and hence the return period analysis. 

\clearpage

\begin{table}[ht]
\centering
\caption{Global average percentage of grid cells with varying distribution modality between the historical and future periods.}
\label{tab:modality_change}
\resizebox{\textwidth}{!}{%
\begin{tabular}{lrrrrr}
\toprule
Experiment & GWL & Unimodal$\rightarrow$Unimodal & Unimodal$\rightarrow$Bimodal & Bimodal$\rightarrow$Unimodal & Bimodal$\rightarrow$Bimodal \\ \midrule
SSP1-2.6   & 1.0 $^\circ$C & 11.01\%          & 2.79\%           & 2.26\%           & 83.94\%         \\
SSP1-2.6   & 2.0$^\circ$C & 10.31\%          & 3.53\%           & 2.45\%           & 83.71\%          \\
SSP2-4.5   & 1.5$^\circ$C & 11.02\%          & 2.78\%           & 2.26\%           & 83.95\%          \\
SSP2-4.5   & 2.0$^\circ$C & 10.24\%          & 3.56\%           & 2.70\%           & 83.50\%          \\
SSP2-4.5   & 3.0$^\circ$C & 8.79\%           & 4.71\%           & 3.08\%           & 83.42\%          \\
SSP2-4.5   & 4.0$^\circ$C & 7.21\%           & 6.02\%           & 3.42\%           & 83.35\%          \\
SSP3-7.0   & 1.5$^\circ$C & 10.82\%          & 2.95\%           & 2.40\%           & 83.83\%          \\
SSP3-7.0   & 2.0$^\circ$C & 10.15\%          & 3.62\%           & 2.89\%           & 83.34\%          \\
SSP3-7.0   & 3.0$^\circ$C & 8.92\%           & 4.68\%           & 3.44\%           & 82.96\%          \\
SSP3-7.0   & 4.0$^\circ$C & 7.80\%           & 5.50\%           & 3.81\%           & 82.89\%          \\
SSP5-8.5   & 1.5$^\circ$C & 11.05\%          & 2.85\%           & 2.31\%           & 83.78\%          \\
SSP5-8.5   & 2.0$^\circ$C & 10.32\%          & 3.58\%           & 3.04\%           & 83.06\%          \\
SSP5-8.5   & 3.0$^\circ$C & 9.14\%           & 4.76\%           & 3.78\%           & 82.32\%          \\
SSP5-8.5   & 4.0$^\circ$C & 8.21\%           & 5.47\%           & 3.88\%           & 82.45\%          \\\bottomrule
\end{tabular}%
}
\end{table}

\begin{figure}[!ht]
    \centering
    \includegraphics[width=\textwidth]{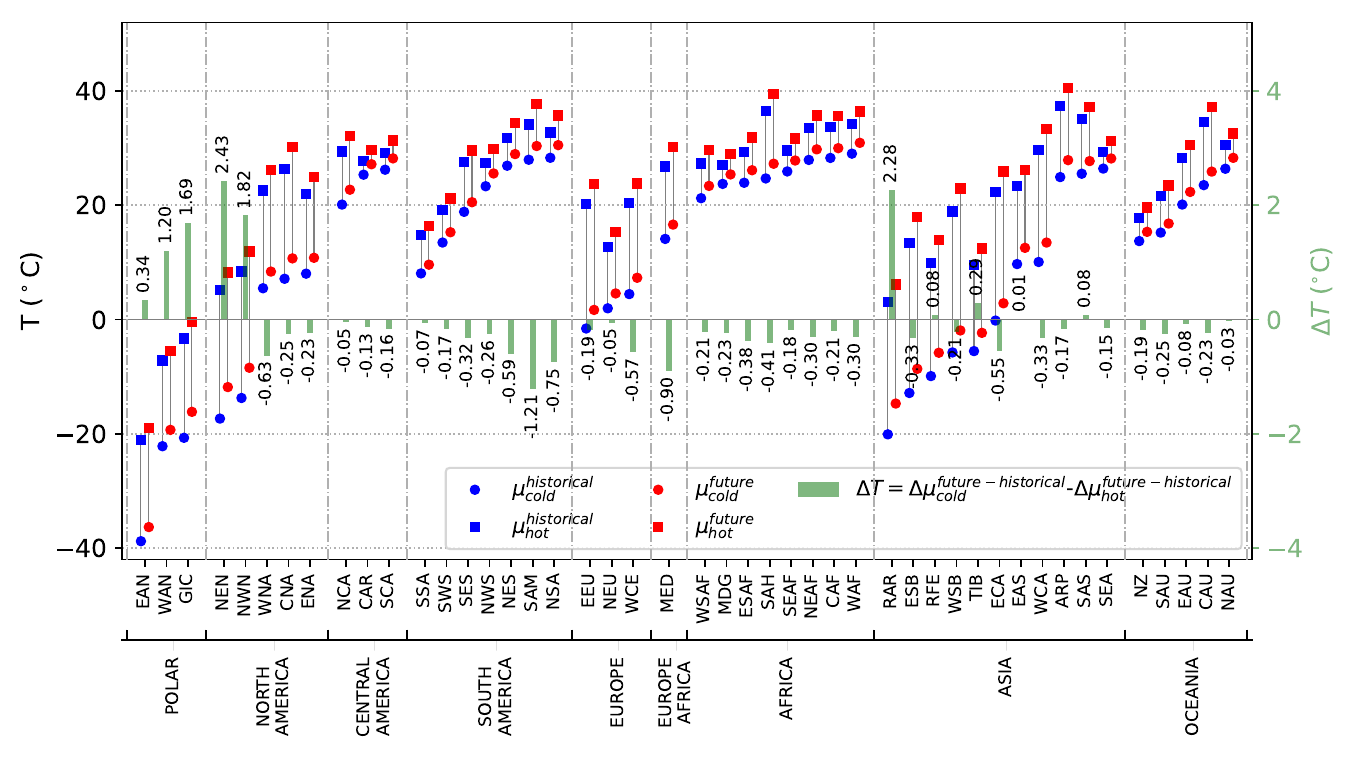}
    \caption{Multi-model peak mean change of region temperature distributions from bimodal grid cells for SSP5-8.5 under GWL3.0$^\circ$C. Blue (red) dots and squares are the means for cold (hot) peaks of the historical (future) period, respectively. They are plotted on the left y-axis. Green bars describe $\Delta T$, the change in the difference between the means of cold and hot Gaussian components, and are plotted on the right y-axis. The upward shift in markers represents the overall warming (see Supplementary Material Figure S7 to S9 for other warming levels).}
    \label{fig:peakchange-585-30}
\end{figure}

\begin{figure}[!ht]
    \centering
    \includegraphics[width=\textwidth]{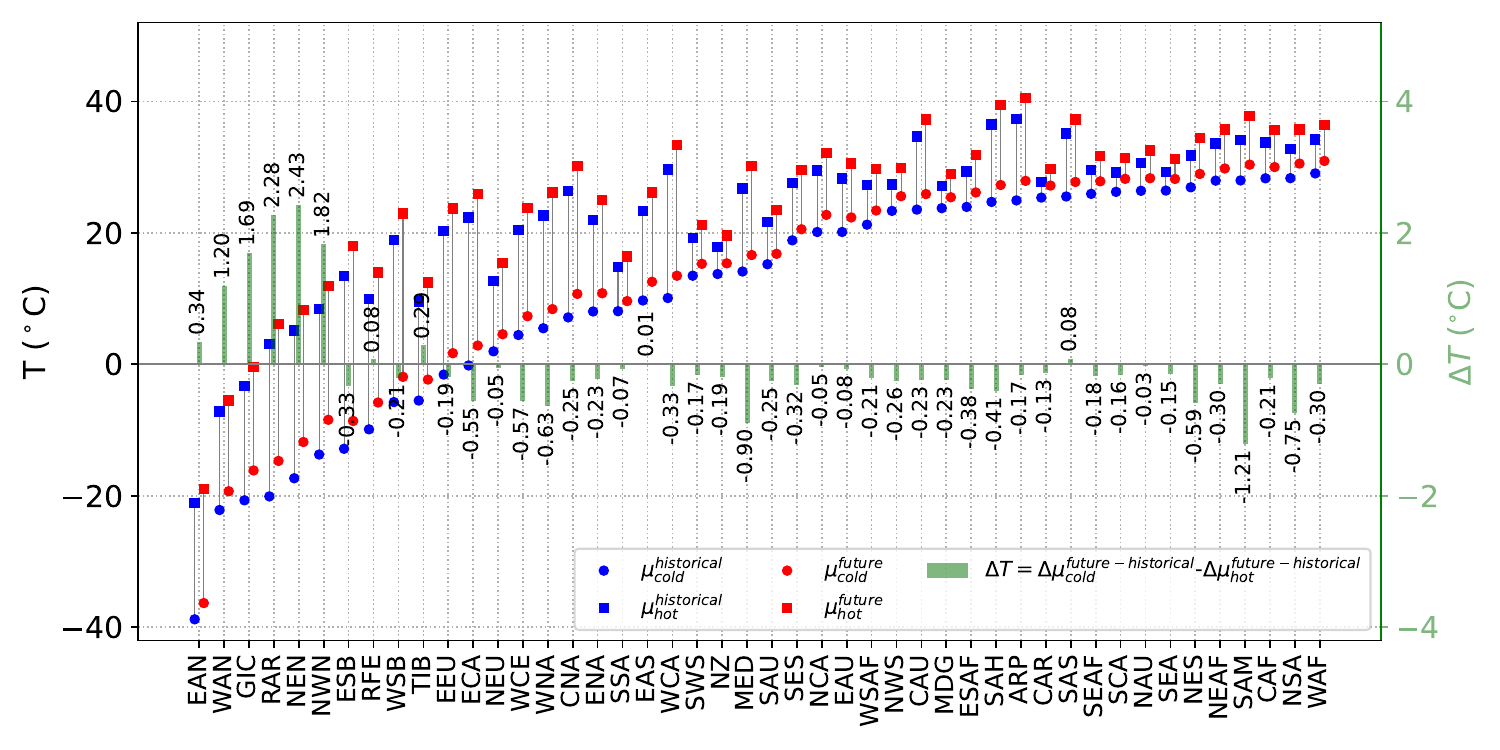}
    \caption{Multi-model peak mean change of region temperature distributions sorted by cold Gaussian mean temperatures (blue dots) for SSP5-8.5 under GWL 3.0$^\circ$C. Blue (red) dots and squares are the means for cold and hot peaks of the historical (future) period, respectively. They are plotted on the left y-axis. Green bars describe $\Delta T$, the change in the difference between the means of cold and hot Gaussian components, and are plotted on the right y-axis. The colder regions have positive $\Delta T$ values and their absolute values are higher than the other regions. The upward shift in blue dots shows that the temperature of cold days is getting warmer and this increase is faster in polar regions compared to the rest of the world (see Supplementary Material Figure S10 to S12 for other warming levels).}
    \label{fig:peakchange-585-30_cold}
\end{figure}

We also analysed the movements of the Gaussian components relative to each other using the $\Delta T$ definition from Equation \ref{eq:deltat} in grid cells with a bimodal distribution. Figure \ref{fig:peakchange-585-30} shows the $\Delta T$ results for all analysed regions for SSP5-8.5 under 3.0$^\circ$C warming (see Supplementary Material Figure S7 to S12 for other warming levels). Changes in distribution peaks are smaller for the lower warming levels. This is consistent with the fact that the time periods for exceeding warming levels are very close to the historical period as shown in Figure \ref{fig:ipcc_regions_data_periods_585}. For the future 3.0$^\circ$C warming scenario, we observed that the mean temperatures are increasing in all regions. Temperature distributions for the European regions have negative $\Delta T$ values, -0.42 degrees on average. This will cause already bimodal peaks in the historical period to separate further from each other in the future, while the whole distribution moves towards higher temperatures. Divergence of peaks will result in more extreme hot temperatures in Europe, as the hot Gaussian moves faster. This result is in agreement with findings from the IPCC AR6 report, in which temperatures in Europe are reported to increase faster than the rest of the globe \cite{IPCC2021}. Polar regions, Northern America and parts of Northern Asia have positive $\Delta T$ values, i.e. converging peaks in grid cells with bimodal distributions. The distribution shape shifts to warmer temperatures and approaches a unimodal distribution as the cold Gaussian part of the distribution moves toward the warmer temperatures faster than the hot Gaussian part. This convergence is also consistent with the slight increase in the percentage of unimodal distribution in polar regions as shown in Figure \ref{fig:modality_58530}. This will cause polar regions to have more days with warmer temperatures also in the colder mode while also having an overall warmer climate. The convergence of peaks in three polar regions (EAN, WAN, GIC) and three northern regions (RAR, NEN and NWN) becomes clear when the regions are sorted by the mean temperature of cold Gaussian component as shown in Figure \ref{fig:peakchange-585-30_cold}. High $\Delta T$ values in polar regions are also supported by previous studies reporting that Arctic regions are warming faster than the global average \cite{Taylor2022}. The lowest $\Delta T$ values are in MED and SAM regions, -0.90 and -1.21 degrees respectively, which will cause both bimodal peaks to diverge from each other while both are moving towards warmer temperatures. Regions in Oceania, Central- and parts of South-America have $\Delta T$ values close to zero, i.e. the cold and hot Gaussian peaks shift toward the warmer temperatures at the same rate. This will cause these regions to have warmer cold and hot periods under future global warming levels compared to the historical period. When all regions are considered, we observe that the extreme temperature events will increase everywhere, as the mean temperatures increase in all regions compared to the historical distributions. The fact that the peaks are converging only in cold climate regions while diverging in other regions shows that shifts in the Gaussian components with respect to each other are essential for extreme temperature event analyses as these changes affect the overall distribution shape and extent. Also, these results are consistent with the change in skewness in temperature distribution as shown in previous studies \cite{Tamarin2020, Skelton2020}. \citeA{Skelton2020} found an abrupt change in skewness in Europe. \citeA{Tamarin2020} found that changes in skewness in winter and summer months will cause cold anomalies in Southern Europe, while warm anomalies intensify in Northeastern Europe. They emphasize the importance of analysing the shape of temperature distributions.  

\begin{sidewaysfigure}
\centering
\includegraphics[width=\textheight,height=0.5\textwidth]{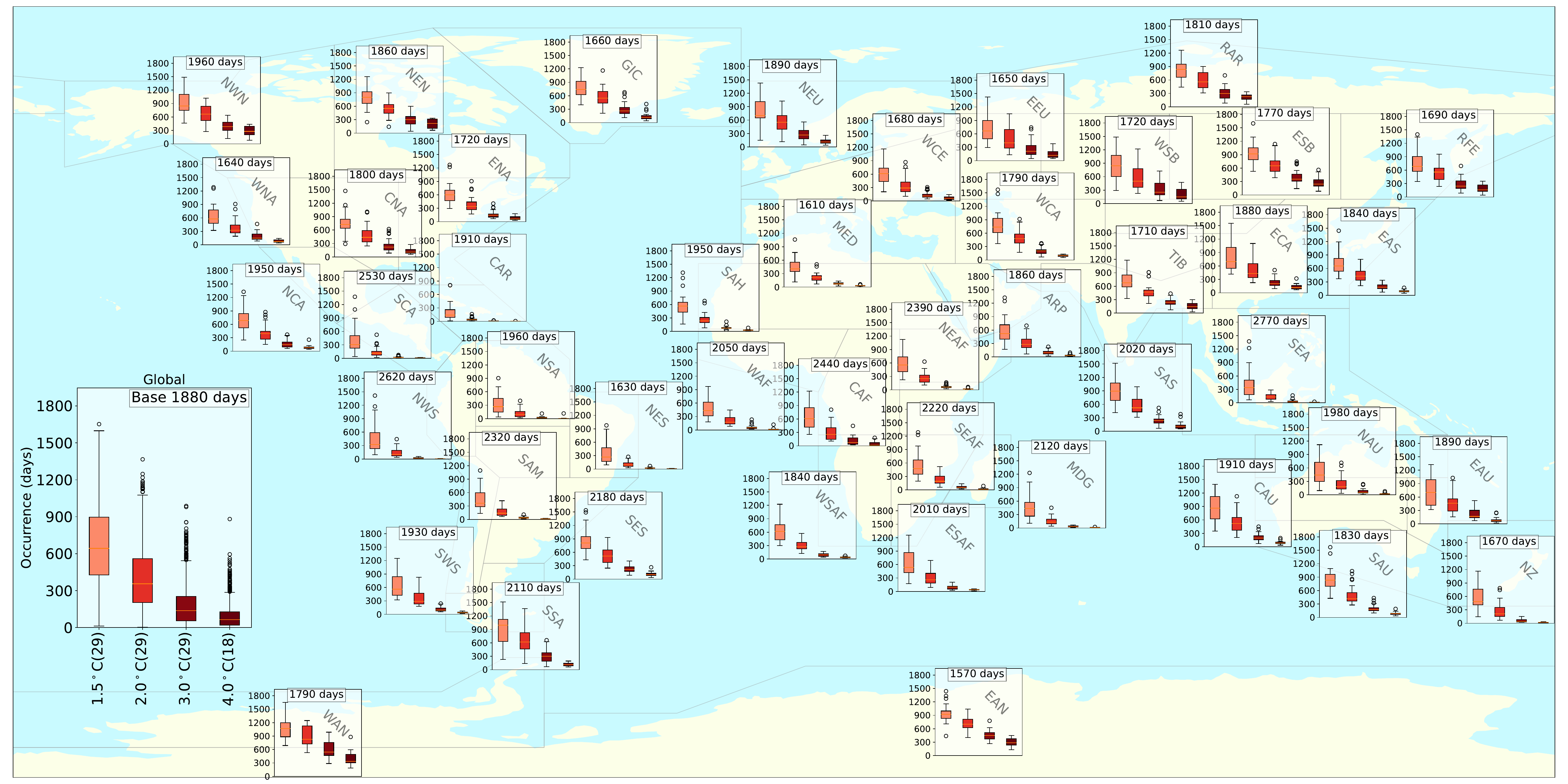}
\caption{Multi-model median of event frequencies for 10-year hot temperature events compared to the 1985-2014 period under GWL 1.5, 2, 3 and 4$^\circ$C relative to 1850-1900 baseline for SSP5-8.5 scenario. The boxes from light to dark shades red represent 1.5$^\circ$C, 2.0$^\circ$C, 3.0$^\circ$C and 4.0$^\circ$C, respectively. The orange lines inside the boxes show the CMIP6 multi-model median, and the boxes extend between the first quartile (Q1) to the third quartile (Q3) of the data, i.e. inter-quartile range (IQR). The vertical lines, i.e. whiskers, stretch out 1.5 IQR from the box. The circles represent the models outside of the interquartile range, i.e. outliers. The length of the hot period used for return period calculations, i.e. number of days in 10 years, is shown in the top right corner of each plot. The global return periods are shown on the left. The more outlier points in the global box plot are because of the differences between regional return periods (See Supplementary Material Figure S13 to S27 for other return periods).}
\label{fig:10yearmap585}
\end{sidewaysfigure}

\begin{figure}[!ht]
    \centering
    \includegraphics[width=0.7\textwidth]{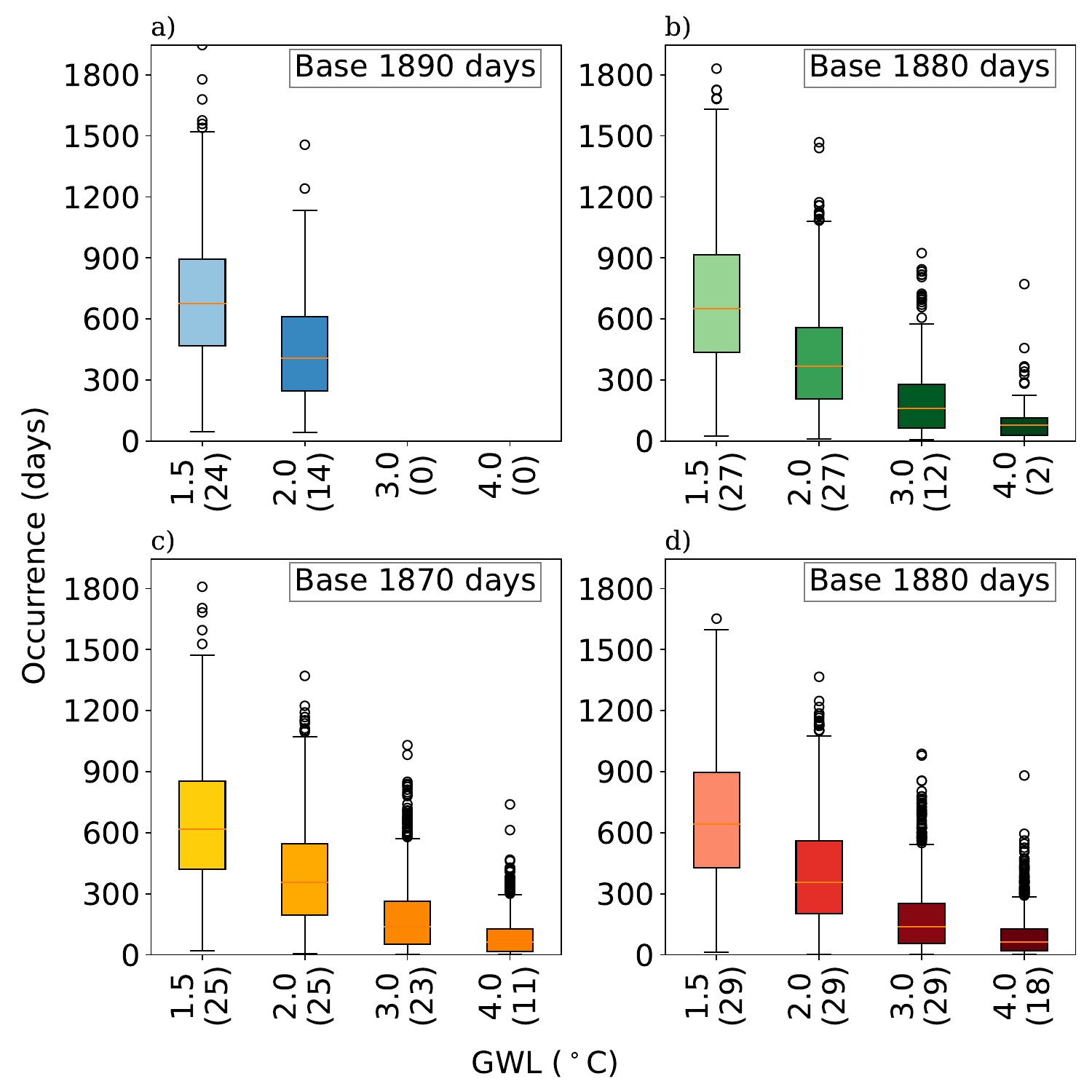}
    \caption{Global multi-model median of event frequencies for 10-year temperature events under 1.5, 2, 3 and 4$^\circ$C warming levels for a) SSP1-2.6, b) SSP2-4.5, c) SSP3-7.0 and d) SSP5-8.5 scenarios. The orange lines inside the boxes show the CMIP6 multi-model median, and the boxes extend between the first quartile (Q1) to the third quartile (Q3) of the data, i.e. inter-quartile range (IQR). The vertical lines, i.e. whiskers, stretch out 1.5 IQR from the box. The circles represent the models outside of the interquartile range, i.e. outliers. The length of the hot period used for return period calculations, i.e. number of days in 10 years, is shown in the top right corner of each plot. The number of datasets is given in parenthesises. All plots show similar results for different SSP scenarios as the GWL are scenario-independent.}
    \label{fig:10year_global}
\end{figure}

After analysing the distribution shapes and peak movements, we calculated the return periods -the average time between the occurrences of a certain event- of 1-year, 5-year, 10-year and 20-year events using only the grid cells with constant modalities, i.e. unimodal or bimodal both for the historical and future periods, as described in Equation \ref{eqn:nyear}. Instead of analyzing extreme temperatures within specific time blocks, our analysis focused on the extremes in the region's probability distribution of 30(20)-years of daily maximum temperatures. Since we used the hottest component in the mixture of Gaussian components to define n-year events, we considered the number of data points falling under the Gaussian component to define \textit{year}-length according to Equation \ref{eqn:yeardef}. For example, globally a 10-\textit{year} event was a temperature event once in every 1880 days (10 years$\textstyle\times 188 \textstyle \frac{days}{year}$) (for bimodal distributions) in the historical period, but it will occur once in every 643, 355, 138, and 63 days under GWL 1.5$^\circ$CC, 2.0$^\circ$C, 3.0$^\circ$C and 4.0$^\circ$C scenarios, as shown in the plot showing global results in Figure \ref{fig:10yearmap585} (also in Figure \ref{fig:10year_global}), respectively. In other words, historical 10-\textit{year} events will be 3.42-\textit{year}, 1.89-\textit{year}, 0.73-\textit{year} and 0.34-\textit{year} events under the future GWL 1.5$^\circ$C, 2.0$^\circ$C, 3.0$^\circ$C and 4.0$^\circ$C scenarios, respectively. After calculating the frequency of extreme events using the temperature distributions in each grid cell individually for an IPCC land region, we averaged the results for the whole region for a single model. The global map with box plots in Figure \ref{fig:10yearmap585} shows multi-model 10-year event frequencies of each region for SSP5-8.5 scenario under different GWL, where the boxes from light to dark shades of red represent 1.5$^\circ$C, 2.0$^\circ$C, 3.0$^\circ$C and 4.0$^\circ$C. Results for 1-year, 5-year, and 20-year events are left out for simplicity and presented in the Supplementary Material Figure S13 to S27. The length of a \textit{``year"} in each region that is used for return period calculations, i.e. the number of days in 10 years, is shown on the top right corner of each sub-plot in Figure \ref{fig:10yearmap585}. 

As shown in Figure \ref{fig:10yearmap585}, return periods of extreme temperature events are getting shorter for all regions under all GWL scenarios as the median of each box is smaller than the historical period. The frequency of extreme events is higher in lower latitudes compared to higher latitudes. For example, the return periods are getting prominently shorter in regions around the equator -where a higher percentage of unimodal grid cells was observed- compared to the other regions. Furthermore, CMIP6 models show narrower boxes and shorter whiskers in lower latitudes compared to wider boxes and longer whiskers in higher latitudes for all analyzed GWL. Among all analysed regions, the Caribbean (CAR) region has the highest increase in the frequency of a 10-year event, from once in 1910 days for the historical period to once in every 137.3, 35.32, 5.5 and 2.0 days under GWL 1.5, 2, 3, and 4$^\circ$C, respectively. Regions around the equator (namely CAR, NSA, NWS, NES, SEA, SCA, SAM, MDG, WAF, and SEAF regions) are the top 10 regions with the highest increase in the frequency of extreme events under all GWL. The frequency of a temperature event equivalent to a 10-year event (historically once in every 1610 days) in the Mediterranean (MED) region increases to once in 405.6, 215.7, 72.4, and 30.6 days in the future under GWL 1.5, 2, 3, and 4$^\circ$C, respectively. Within the European continent, the West\&Central Europe (WCE) region has a higher increase in the frequency of extreme events compared to the Eastern Europe (EEU) and the North Eastern Europe (NEU) regions, where the latter two regions are among the regions with the least increase in extreme temperature event frequency. The smallest increase in the frequency of hot extremes is observed in the Western Antarctica (WAN) region, where the return periods of 10-year events will decrease from once in 1790 days to once in 1070.1, 827.6, 542.7 and 338.7 days under GWL 1.5, 2, 3, and 4$^\circ$C, respectively. High latitude regions, such as WAN, NEU, EAN, NWN, ESB, GIC, RAR, SSA, TIB, and NEN regions are the 10 regions with the smallest decrease in return periods of extreme hot temperature events. Some of these regions are polar regions with positive $\Delta T$ values as shown in Figure \ref{fig:peakchange-585-30_cold}. This will cause more days with warmer temperatures in the colder mode of these regions while having an increase in hot extremes. 

\section{Summary and Discussion} \label{sec:discussion}

Detection of extreme events is important to mitigate their impact on natural and anthropogenic systems. Future projections suggest that the mean and standard deviations of maximum surface temperature will increase. This change in the shape of maximum surface temperature distributions increases the intensity and frequency of extreme events in the future. However, not only the shift to warmer temperatures but also the modality of temperature distribution affects the parameters of the entire distribution which is important to calculate the return periods as shown in this study. 

GMM are a promising method for calculating the return periods of extreme events, and additionally determining the shape of the entire distribution for daily maximum temperature data. GMM can provide information on different climate features in different regions such as cold and hot periods, and their changes. We showed that bimodality is a prominent characteristic observed in multi-year daily near-surface maximum temperature data. To understand the underlying factors of this bimodal pattern, we analyzed temperature distributions from grid cells with distinct bimodality across different months, seasons and 6-month running windows. We observed that the winter and summer seasons emerged as the primary contributors to the peaks observed in the bimodal distribution. In grid cells of different regions with distinct bimodal distributions, the transition from winter to summer occurs swiftly, leading to a more distinct separation of the temperature modes. Consequently, the distributions during transitional seasons, such as spring and autumn, appeared to be wider (covering a broader value range) compared to the more distinct distributions observed during winter and summer (covering a very small value range). Furthermore, analyses of 6-month running windows also showed an agglomeration of similar temperatures around winter(summer) months from November(May) to April(October) that creates the peaks in the bimodal distribution (See Supplementary Material Figure S2 for the distributions of seasons and months.). Here, the advantage of GMM becomes evident. For analyses to uncover the origins of bimodality, we had to select certain seasons or months. Seasonal periods are commonly used in previous studies to analyse extreme events \cite{Qian2015, Qian2019, Walt2021, Prodhomme2022}. For example, \citeA{Qian2015} found that the seasonality is weakening in the northern high-latitude regions and East Asia while strengthening in the Mediterranean. This can also be seen in Figure \ref{fig:peakchange-585-30_cold}, as the northern regions and east/central Asia regions have converging peaks which means that these regions will have a distribution closer to a unimodal distribution. Meanwhile, diverging peaks in MED will introduce more distinct cold and warm periods. However, onsets and length of seasons are predicted to change with climate change \cite{Wang2021}. Therefore, the definition of current seasonal periods or months will not necessarily be valid for future climates. One can utilize GMM to determine the hot Gaussian component of a region to define the length of the analysis period instead of using fixed seasonal definitions. Moreover, the bimodality analysis also shows how peaks are changing in the future, effectively changing the expected climate of the area.

ETCCDI indices are commonly used in extreme event analysis as they offer a simple and concise way to define extremes \cite{Zhang2011, Zhao2021, Vogel2020}. ETCCDI indices use block maxima methods such as TXx (Monthly maximum value of daily max temperature), TNx (Monthly maximum value of daily min temperature) or percentile-over-threshold (POT) methods such as TX90p (Percentage of time when daily max temperature $>90^{th}$percentile), TN90p (Percentage of time when daily min temperature $<90^{th}$percentile) \cite{Zhang2011}. These exceedances can be modelled with GEV distributions or generalized Pareto distribution (GPD). However, GEV distributions are a better fit for longer block sizes than for shorter blocks like daily data. If the available dataset is short, the longer block sizes will produce fewer data which can increase the variability in parameter estimation \cite{Huang2016, Wang2016}. For example, if there is more than one extremely hot day in the block (month, season or year), e.g. several consecutive days, block maxima methods consider only the hottest, and hence only one day in a block, while GMM considers all days hotter than the threshold. Assuming that a heat wave lasts usually days to a few weeks, a substantial number of hot days might not be seen by block maxima methods as long as they fall into the same block. Percentile-over-threshold methods together with count-day indices such as WSDI (Warm spell duration indicator) are useful for analysing the durations of events. However, the derivation of percentiles is strongly affected by the choice of the base period, a right shift in the distribution will result in a higher threshold and erroneously reduce the frequency of extreme events \cite{Yosef2021}. Seasonality in temperature extremes adds complexity to the process of selecting percentiles to define extreme temperatures \cite{Huang2016}. The advantage of GMM is that the model analyses the distribution of temperatures without any previous assumption and learns the hot periods from the data. Also,  GMM uses all available data in contrast to block maxima methods, which makes it useful if the available data is short or bimodality exists \cite{Sardeshmukh2015, Wang2016, Knoben2019, Alaya2020}.

However, since the Gaussian components of GMM are not bounded, it is important to only calculate the return periods of extreme events equal to or less than the study period when applying GMM. Additionally, we only used grid cells which have the same number of Gaussian components in their temperature distribution, i.e. unimodal or bimodal distribution, both for the historical and future periods. Grid cells with changing distribution shapes, e.g. transforming from a bimodal distribution in the historical period to a unimodal distribution in the future or vice versa, were found in less than 10\% of the grid cell for each GWL as shown in Table \ref{tab:modality_change}, and were disregarded in the analysis as calculating the temperature thresholds becomes problematic with the abrupt change in means and standard deviations. 

For the first time, the IPCC AR6 Report includes a new dedicated chapter on weather and climate extreme events \cite{IPCC2021}. This emphasizes the importance of robust methods of extreme event detection to be able to mitigate the impact of such events. IPCC AR6 reports that the return periods of 10-year events will increase around the world, with the highest changes projected to happen in some mid-latitude and semi-arid regions. Our findings are in agreement with these results. Furthermore, IPCC AR6 projects the warming rate in mid-latitudes to be higher than the average global warming rate. GMM might explain why these regions are projected to have higher warming, as we observed that grid cells in these regions predominantly follow a bimodal distribution in the historical (future) period as shown in Figure \ref{fig:modality_hist} (\ref{fig:modality_58530}). Furthermore, these regions have diverging peaks as shown in Figure \ref{fig:peakchange-585-30_cold}, i.e. mode for warm temperatures moving towards warmer temperatures faster than the mode for colder temperatures. These diverging bimodal peaks will create distinct Gaussian components in the entire multi-year daily maximum temperature, which in turn results in a higher increase in extremes in these regions. For example, almost all grid cells in the Mediterranean region follow a bimodal distribution, and the peaks of bimodal distribution will diverge in the future as shown in Figure \ref{fig:peakchange-585-30}. Mediterranean region is identified as one of the most responsive regions to climate change and a hot spot of climate extremes \cite{IPCC2021, Feng2022}. Similarly, Arctic regions are projected to have the highest increase in temperature of the coldest days \cite{IPCC2021, Li2021}. Our results are also consistent with these increases as shown in Figure \ref{fig:peakchange-585-30}, where diverging bimodal peaks in mid-latitude regions will shift the mode for warm temperatures, i.e. hot Gaussian, to the higher temperature ranges. This shift in the Gaussian components of temperature distribution will cause those land regions to have warmer temperature extremes and can explain the higher average warming rate than the global average. Likewise, converging peaks in polar regions as shown in Figure \ref{fig:peakchange-585-30} will move the cold Gaussian part toward warmer temperatures, thereby introducing higher warming on the coldest days.

According to our analyses, 10-year events will increase almost 3-fold under GWL 1.5$^\circ$C compared to the historical period for all SSP scenarios as shown in Figure \ref{fig:10year_global} when looking at the whole globe. This means a temperature event that occurs once in every 10 years (1880 days) will be expected to occur 2.9 times in every 10 years under GWL 1.5$^\circ$C. 10-year extreme temperature events will become even more frequent globally under GWL 2$^\circ$C, 3$^\circ$C and 4$^\circ$C; 5.3, 13.6, and 29.5 times every 10 years, respectively. In other words, current 10-year events will be 3.42-\textit{year}, 1.89-\textit{year}, 0.73-\textit{year} and 0.34-\textit{year} events in the future under GWL 1.5$^\circ$C, 2$^\circ$C, 3$^\circ$C and 4$^\circ$C, respectively. Our results show a higher increase compared to the IPCC AR6 report, where the frequency of 10-year events is projected to increase approximately 3, 4, 5.5 and 9-fold under GWL 1.5$^\circ$C, 2$^\circ$C, 3$^\circ$C and 4$^\circ$C, respectively \cite{IPCC2021}, using a block maxima method for determining the extreme events. The higher increase in our method compared to IPCC AR6 can most likely be explained by the fact that we used GMM to model the distribution of temperatures and GMM considers all days hotter than the threshold, while the block maxima method only uses the maximum of a block. Another important point deduced from the analyses of different regions for several CMIP6 models is that the ensemble of analyzed CMIP6 models shows coherent results for regions as shown in the regional box plots in Figure \ref{fig:10yearmap585}. Most of the individual model results fall within the first and third quartile, and only a few models fall outside this range. The higher number of outlier points in the global box plot in Figure \ref{fig:10yearmap585}, and also shown for different SSP scenarios in Figure \ref{fig:10year_global}, are caused by the differences between regional return periods. All SSP scenarios show similar results with each other as the return periods are calculated for GWL which have the same forcing on climate.

Return periods of extreme events become shorter in every region, which means that the frequency of extreme temperature events increases. This will become larger with increasing global warming levels. Some climate models have already exceeded GWL 1.5$^\circ$C with respect to the 1850-1900 period as shown in Figure \ref{fig:ipcc_regions_data_periods_585}. This fact further emphasises the importance of robust methods to detect extreme events. Even though there is a delay in taking the necessary precautions to reduce the speed of the warming of the climate, as time goes by, tomorrow's projections become today's reality.

\section*{Code and data availability}
The recipes to extract regional data from CMIP6 models using ESMValTool, python scripts to analyse extreme events and to produce all figures of this manuscript are accessible in the following GitHub repository: \url{https://github.com/EyringMLClimateGroup/pacal23jgr_GaussianMixtureModels_Extremes}. The regional output files amount to hundreds of GB.

The latest release of ESMValTool is publicly at \url{https://github.com/ESMValGroup/ESMValTool} \cite{Andela_ESMValTool_2022}.

\acknowledgments
Funding for this study was provided by the European Research Council (ERC) Synergy Grant “Understanding and Modelling the Earth System with Machine Learning (USMILE)” under the Horizon 2020 research and innovation programme (Grant Agreement No. 855187). This work used resources of the Deutsches Klimarechenzentrum (DKRZ) granted by its Scientific Steering Committee (WLA) under project ID bd1083. We acknowledge the World Climate Research Programme, which, through its Working Group on Coupled Modelling, coordinated and promoted CMIP6. We thank the climate modelling groups for producing and making available their model outputs, the Earth System Grid Federation (ESGF) for archiving the data and providing access, and the multiple funding agencies that support CMIP6 and ESGF. \citeA{ERA5_data} was downloaded from the \citeA{C3S}. The results contain modified Copernicus Climate Change Service information 2020. Neither the European Commission nor ECMWF is responsible for any use that may be made of the Copernicus information or data it contains. We would like to thank Dr. Pauline Bonnet for her valuable comments and suggestions to improve the manuscript. We would like to extend our sincere gratitude to the anonymous reviewers whose invaluable feedback and constructive comments significantly contributed to the improvement and quality of this work.

\bibliography{bibliography}

\end{document}


\title{Supporting Information for "Detecting Extreme Temperature Events Using Gaussian Mixture Models"}

\authors{Aytaç Paçal\affil{1,2}, Birgit Hassler\affil{1}, Katja Weigel\affil{2,1}, M. Levent Kurnaz \affil{4}, Michael F. Wehner\affil{3}, Veronika Eyring\affil{1,2}}

\affiliation{1}{Deutsches Zentrum für Luft- und Raumfahrt e.V. (DLR), Institut für Physik der Atmosphäre, Oberpfaffenhofen, Germany}
\affiliation{2}{University of Bremen, Institute of Environmental Physics (IUP), Bremen, Germany}
\affiliation{3}{Computational Research Division, Lawrence Berkeley National Laboratory, Berkeley, CA, USA}
\affiliation{4}{Center for Climate Change and Policy Studies, Boğaziçi University, Istanbul, Turkey}

\begin{article}

\begin{figure}
    \centering
    \includegraphics[width=\textwidth]{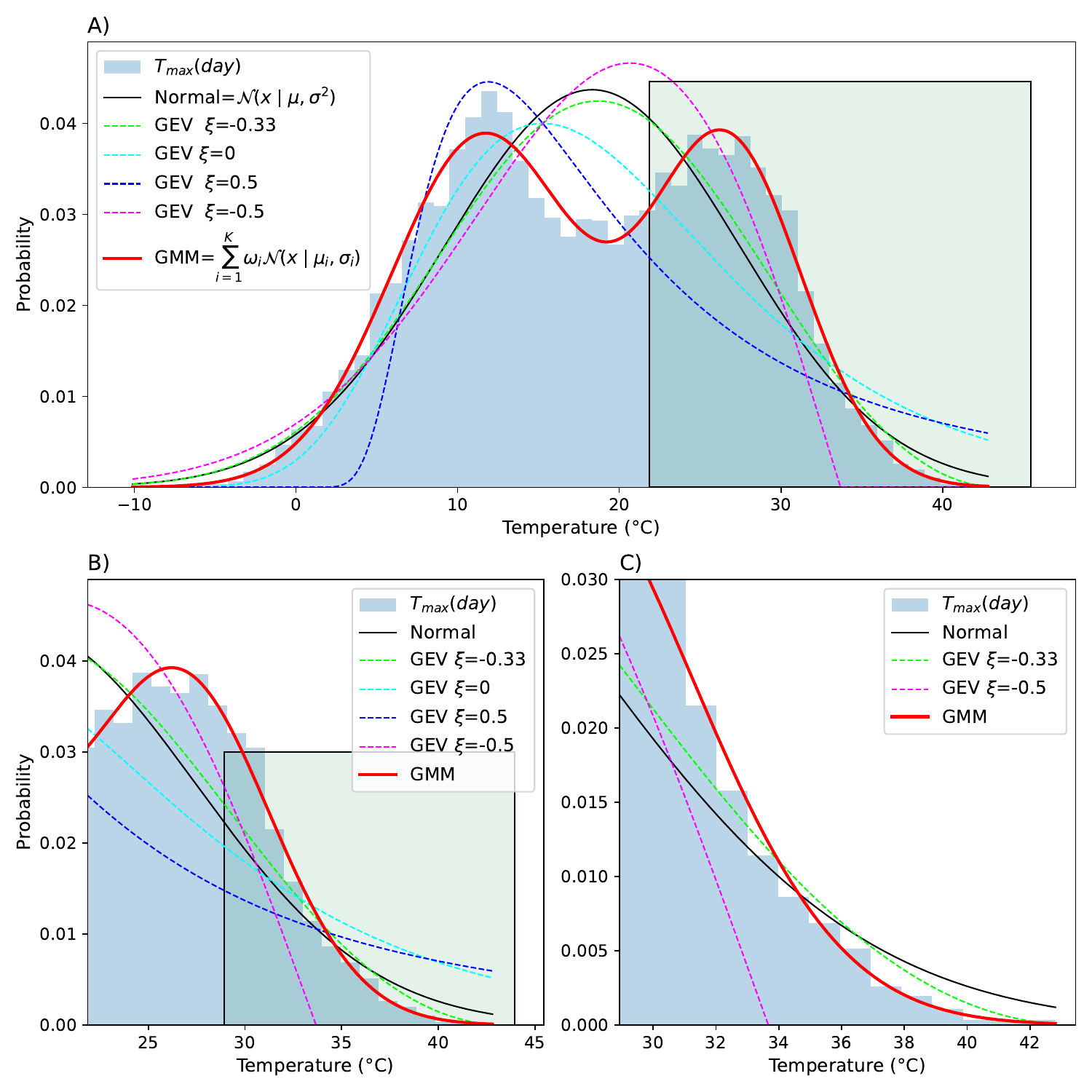}
    \caption{A) Comparison of the GMM, GEV and Normal distributions for the maximum daily temperature, B) and C) zoomed on the tails}
    \label{fig:gmm_comparison}
\end{figure}

\begin{figure}
    \centering
    \includegraphics[width=\textwidth]{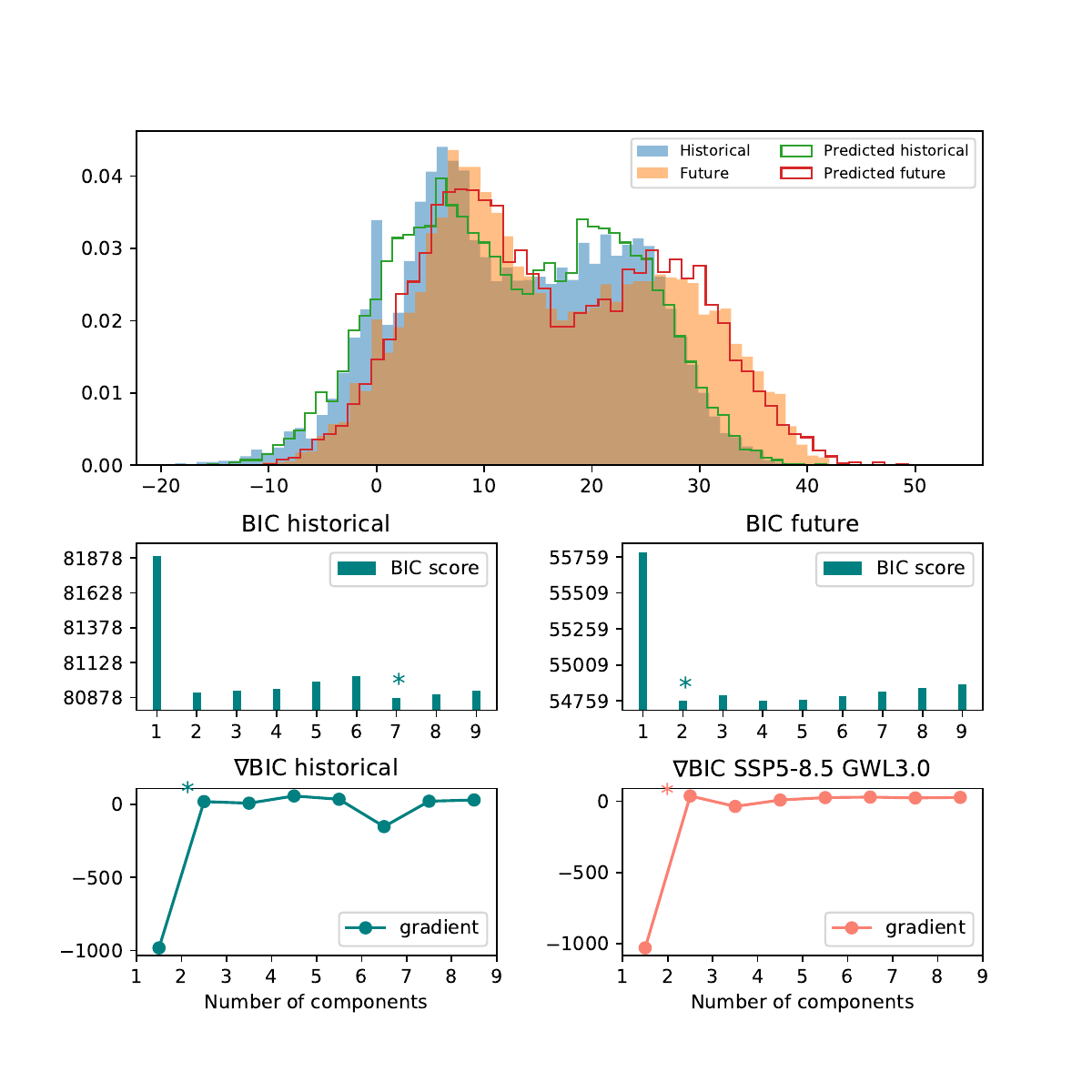}
    \caption{Choosing the optimal number of components. (Top) Blue and orange bars represent multi-year raw daily maximum temperature data from a grid cell (Lat=43.48, Lon=19.69) in the MED region from the MPI-ESM1-2-HR model for the 1985-2014 historical and SSP5-8.5 GWL3$^\circ$C future period. Green and red lines represent the predicted GMM for historical and future periods. (Middle) The left and right plots show BIC scores for the number of GMM components for historical and future data, respectively. Asterisks show the chosen number of components with the lowest BIC score. (Bottom) The left and right plots show the highest change in BIC scores for historical and future periods, respectively. Asterisks show the chosen number of components after the highest gradient and the unimodality check.}
    \label{fig:gmm_selection}
\end{figure}

\begin{figure}[ht]
    \centering
    \includegraphics[width=0.9\textwidth]{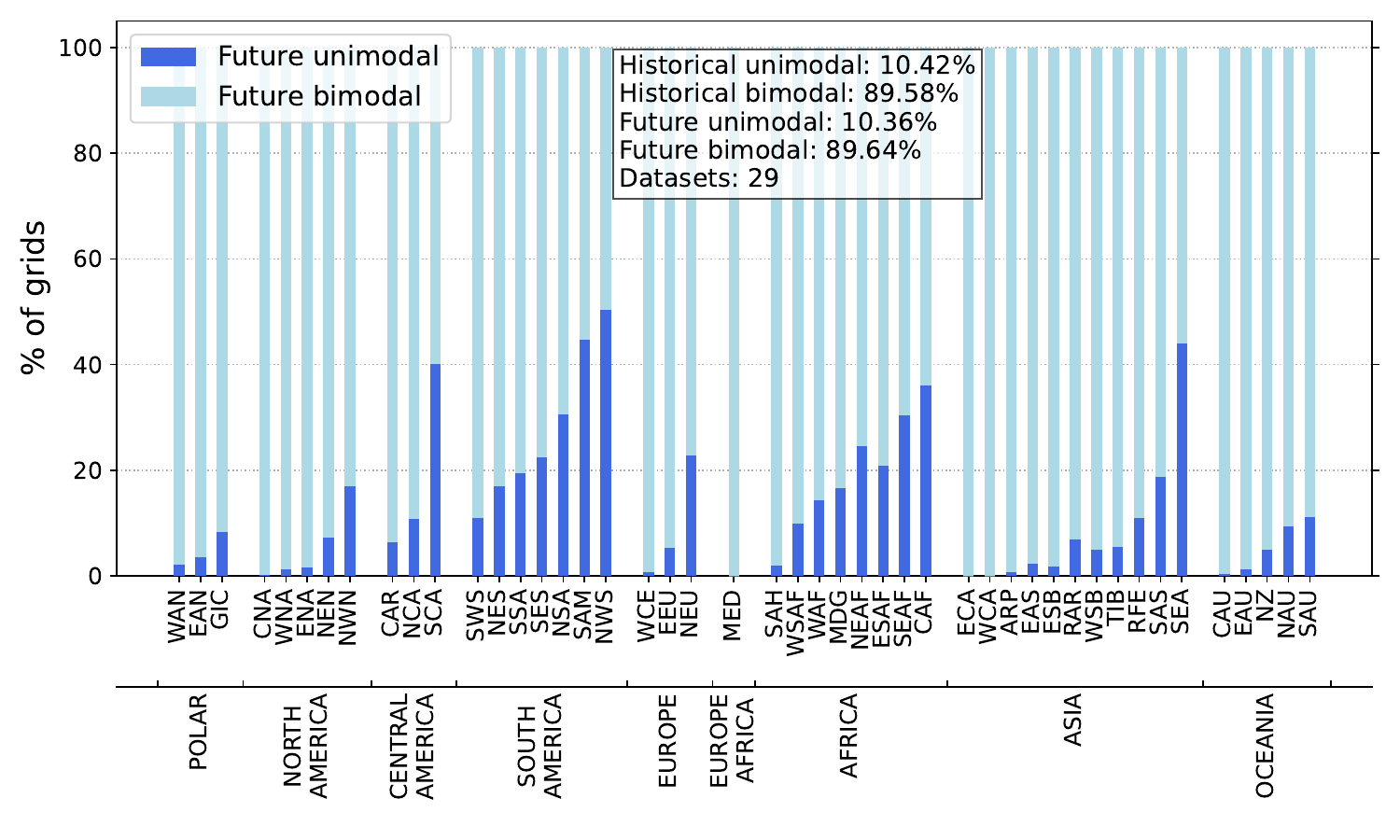}
    \caption{Multi-model mean percentages of grid modalities for future period in study regions grouped by continents under GWL 1.5$^\circ$C for SSP5-8.5 scenario}
    \label{fig:modality_58515}
\end{figure}

\begin{figure}[ht]
    \centering
    \includegraphics[width=0.9\textwidth]{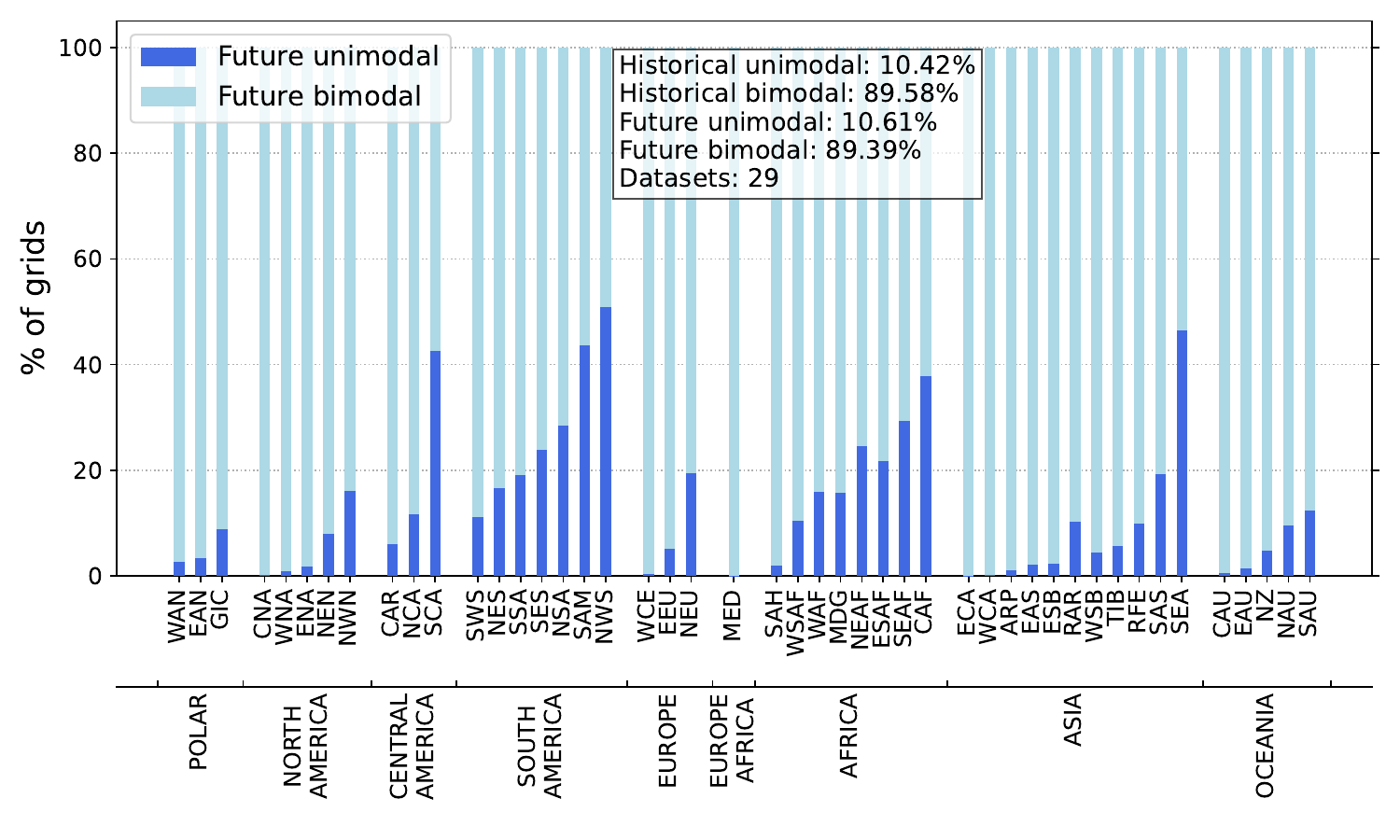}
    \caption{Same as Figure \ref{fig:modality_58515} but for GWL 2.0$^\circ$C for SSP5-8.5 scenario}
    \label{fig:modality_58520}
\end{figure}

\begin{figure}[ht]
    \centering
    \includegraphics[width=0.9\textwidth]{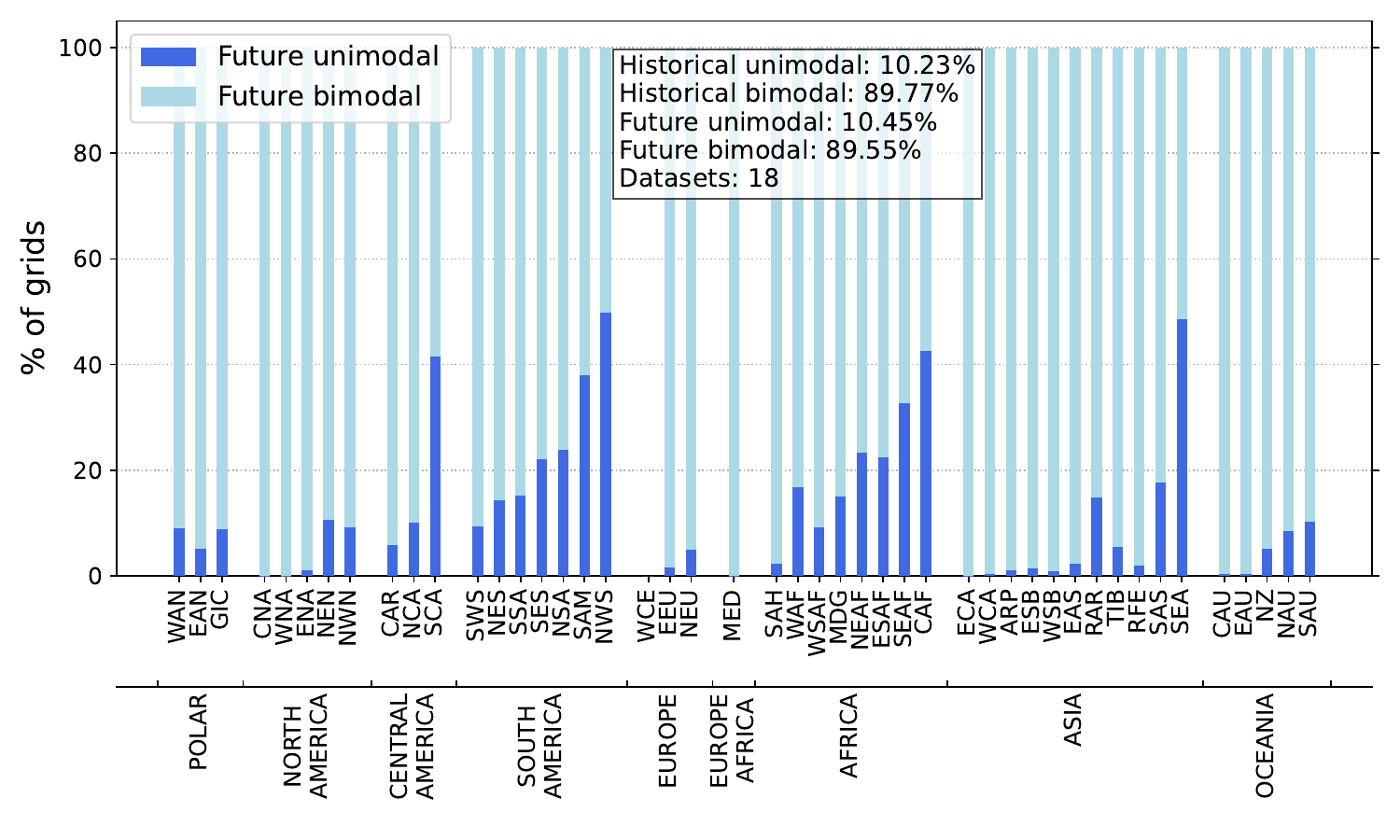}
    \caption{Same as Figure \ref{fig:modality_58515} but for GWL 4.0$^\circ$C for SSP5-8.5 scenario}
    \label{fig:modality_58540}
\end{figure}

\begin{figure}[ht]
    \centering
    \includegraphics[width=0.9\textwidth]{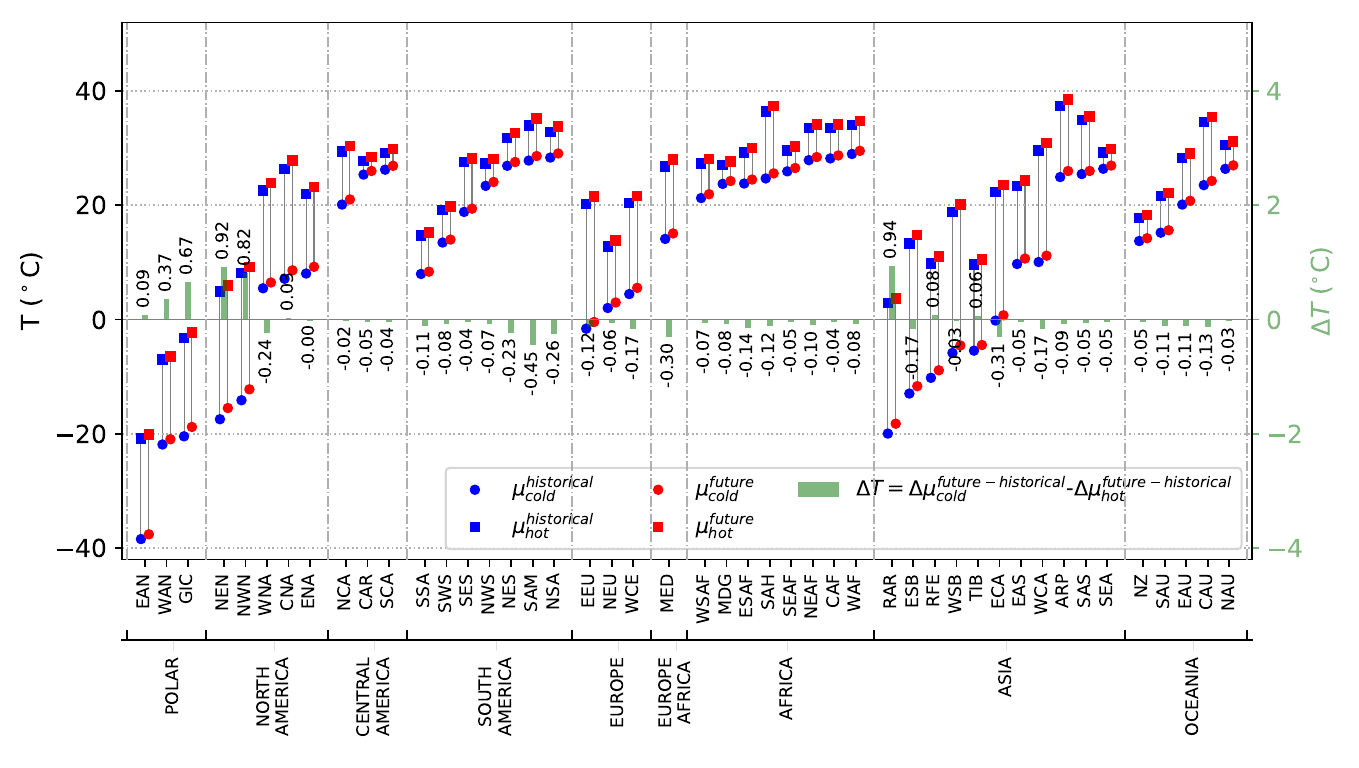}
    \caption{Multi-model mean change of region temperature distributions under GWL 1.5$^\circ$C for SSP5-8.5 scenario. Blue (red) dots and squares are the means for cold and hot peaks of the historical (future) period, respectively. They are plotted on the left y-axis. Green bars describe $\Delta T$, the change in the difference in differences between the means of cold and hot Gaussian components, and are plotted on the right y-axis. The upward shift in markers represents the overall warming.}
    \label{fig:peakchange_58515}
\end{figure}

\begin{figure}[ht]
    \centering
    \includegraphics[width=0.9\textwidth]{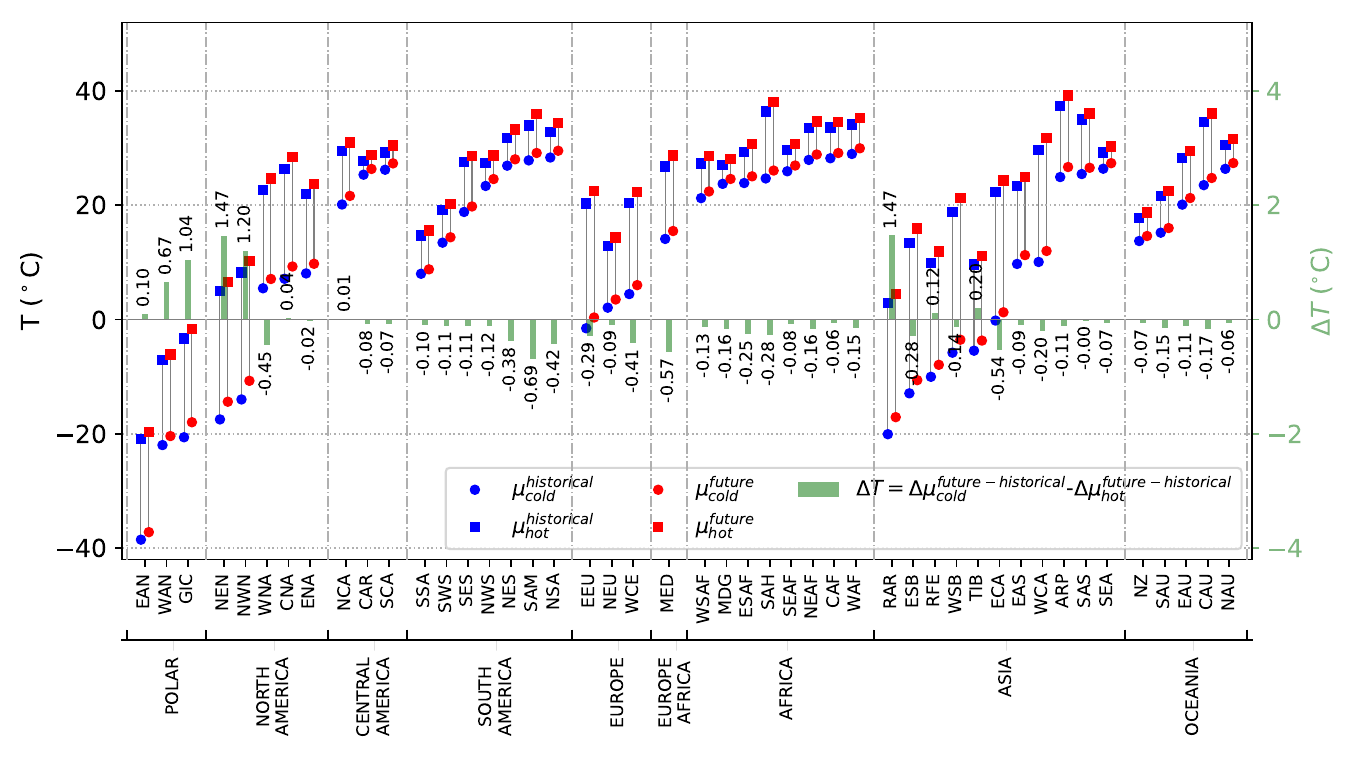}
    \caption{Same as Figure \ref{fig:peakchange_58515} but for GWL 2.0$^\circ$C for SSP5-8.5 scenario.}
    \label{fig:peakchange_58520}
\end{figure}

\begin{figure}[ht]
    \centering
    \includegraphics[width=0.9\textwidth]{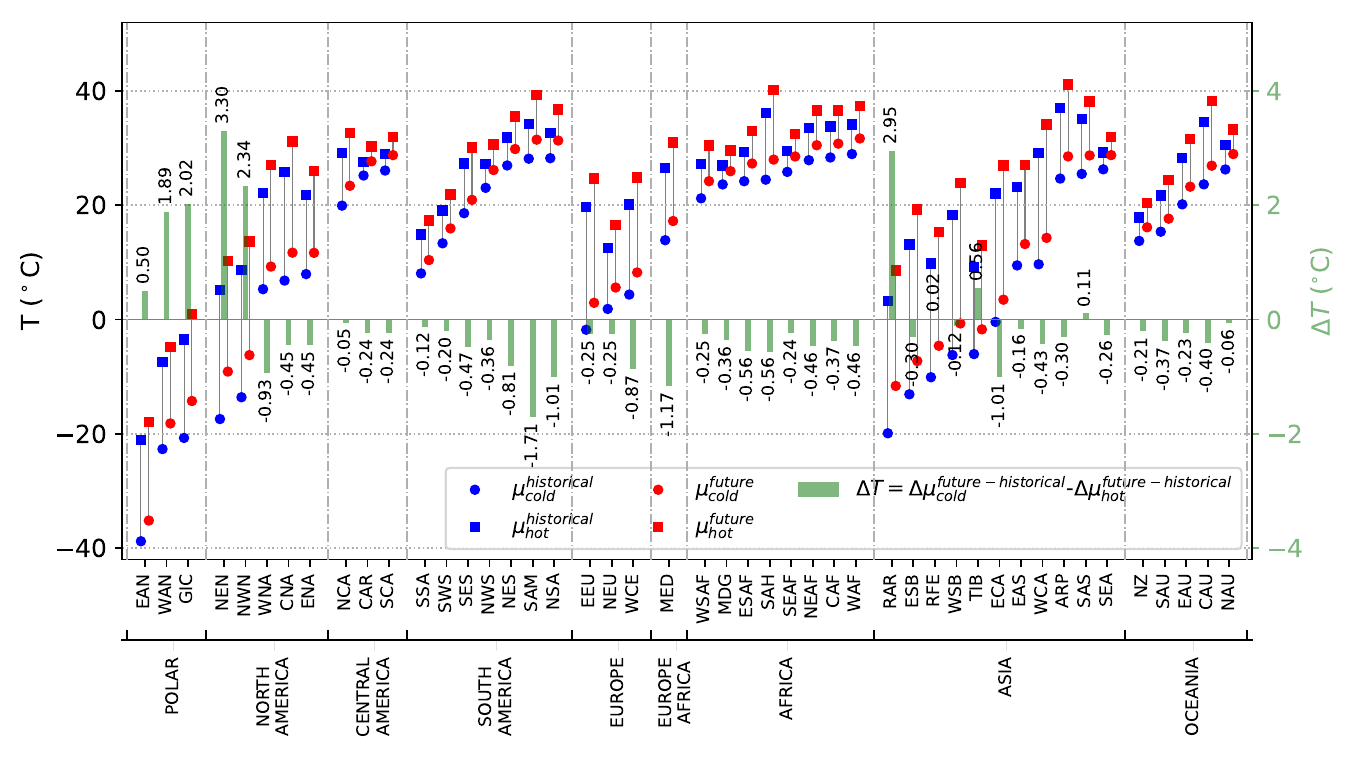}
    \caption{Same as Figure \ref{fig:peakchange_58515} but for GWL 4.0$^\circ$C for SSP5-8.5 scenario.}
    \label{fig:peakchange_58540}
\end{figure}

\begin{figure}[ht]
    \centering
    \includegraphics[width=\textwidth]{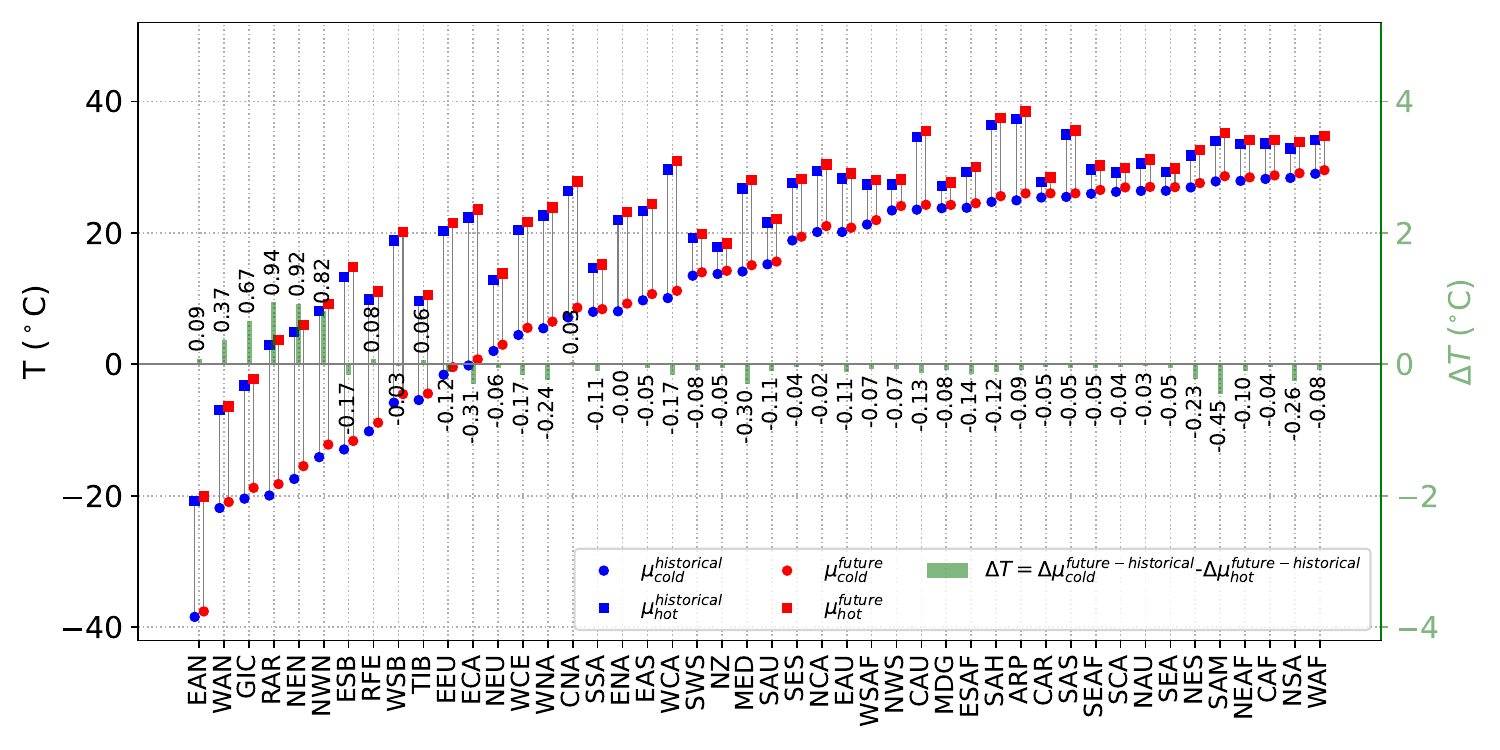}
    \caption{Multi-model mean change of region temperature distributions under GWL 1.5$^\circ$C for SSP5-8.5 ordered by mean temperature of region's cold peak. Blue (red) dots and squares are the means for cold and hot peaks of the historical (future) period, respectively. They are plotted on the left y-axis. Green bars describe $\Delta T$, the change in the difference in differences between the means of cold and hot Gaussian components, and are plotted on the right y-axis. The upward shift in markers represents the overall warming.}
    \label{fig:peakchange_ordered_58515}
\end{figure}

\begin{figure}[ht]
    \centering
    \includegraphics[width=\textwidth]{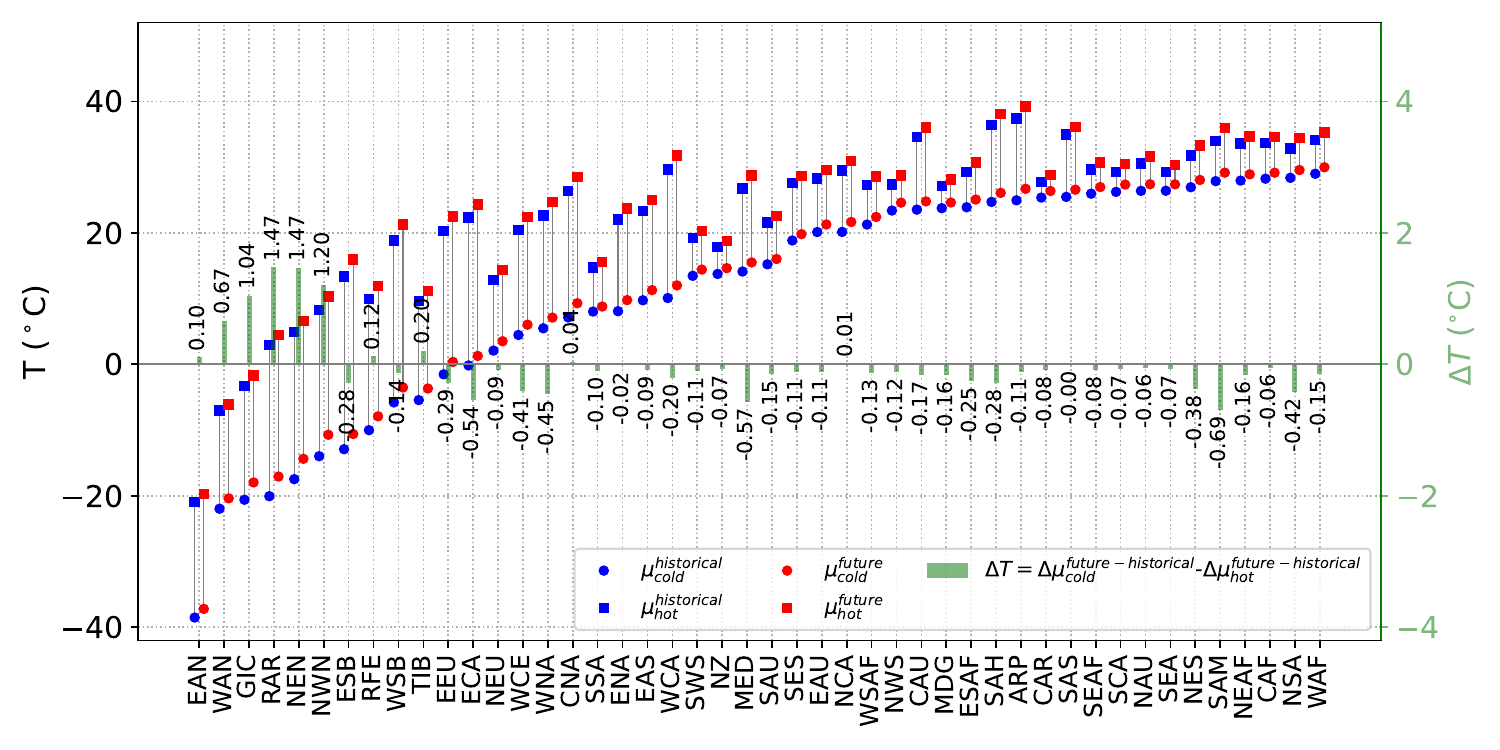}
    \caption{Same as Figure \ref{fig:peakchange_ordered_58515} but for GWL 2.0$^\circ$C for SSP5-8.5 scenario.}
    \label{fig:peakchange_ordered_58520}
\end{figure}

\begin{figure}[ht]
    \centering
    \includegraphics[width=\textwidth]{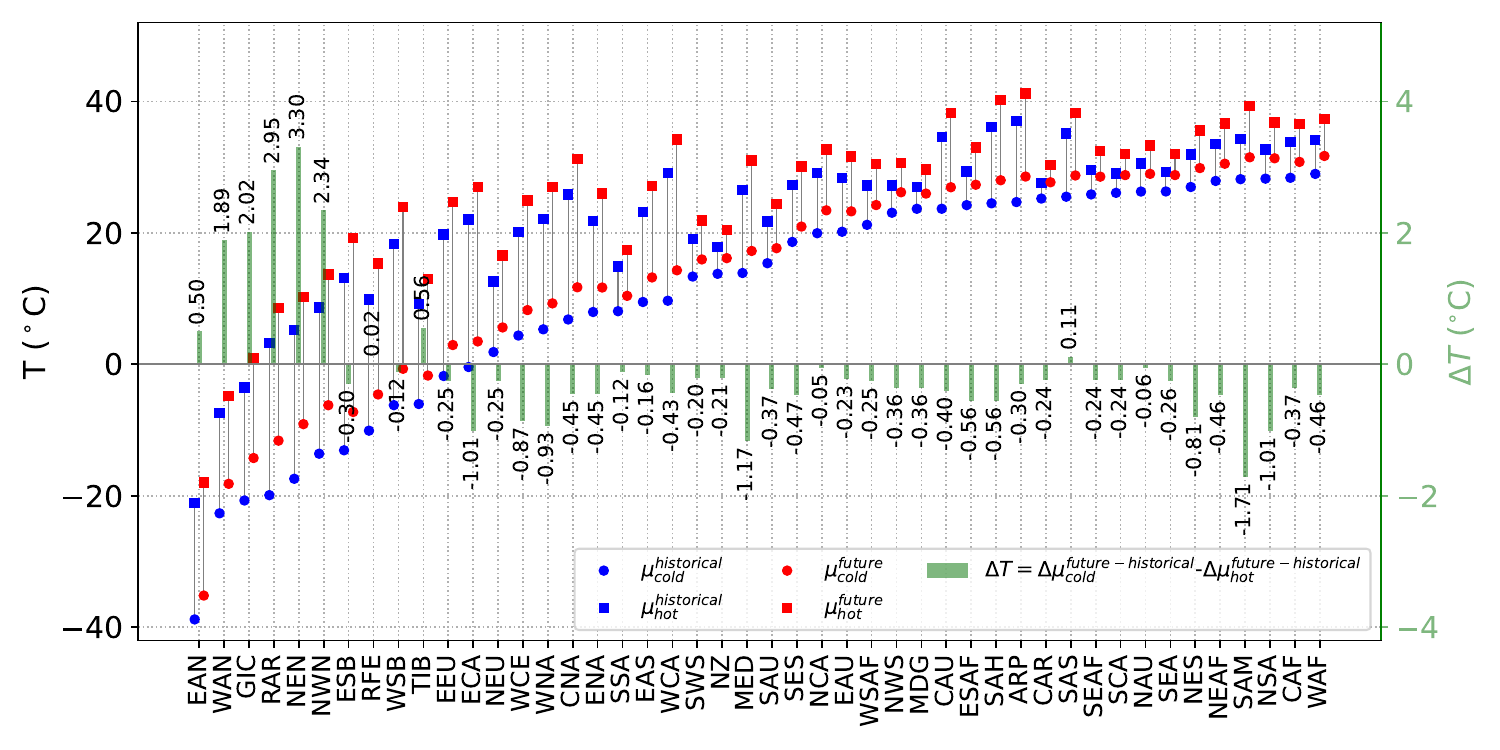}
    \caption{Same as Figure \ref{fig:peakchange_ordered_58515} but for GWL 4.0$^\circ$C for SSP5-8.5 scenario.}
    \label{fig:peakchange_ordered_58540}
\end{figure}

\begin{figure}[ht]
    \centering
    \includegraphics[width=\textwidth]{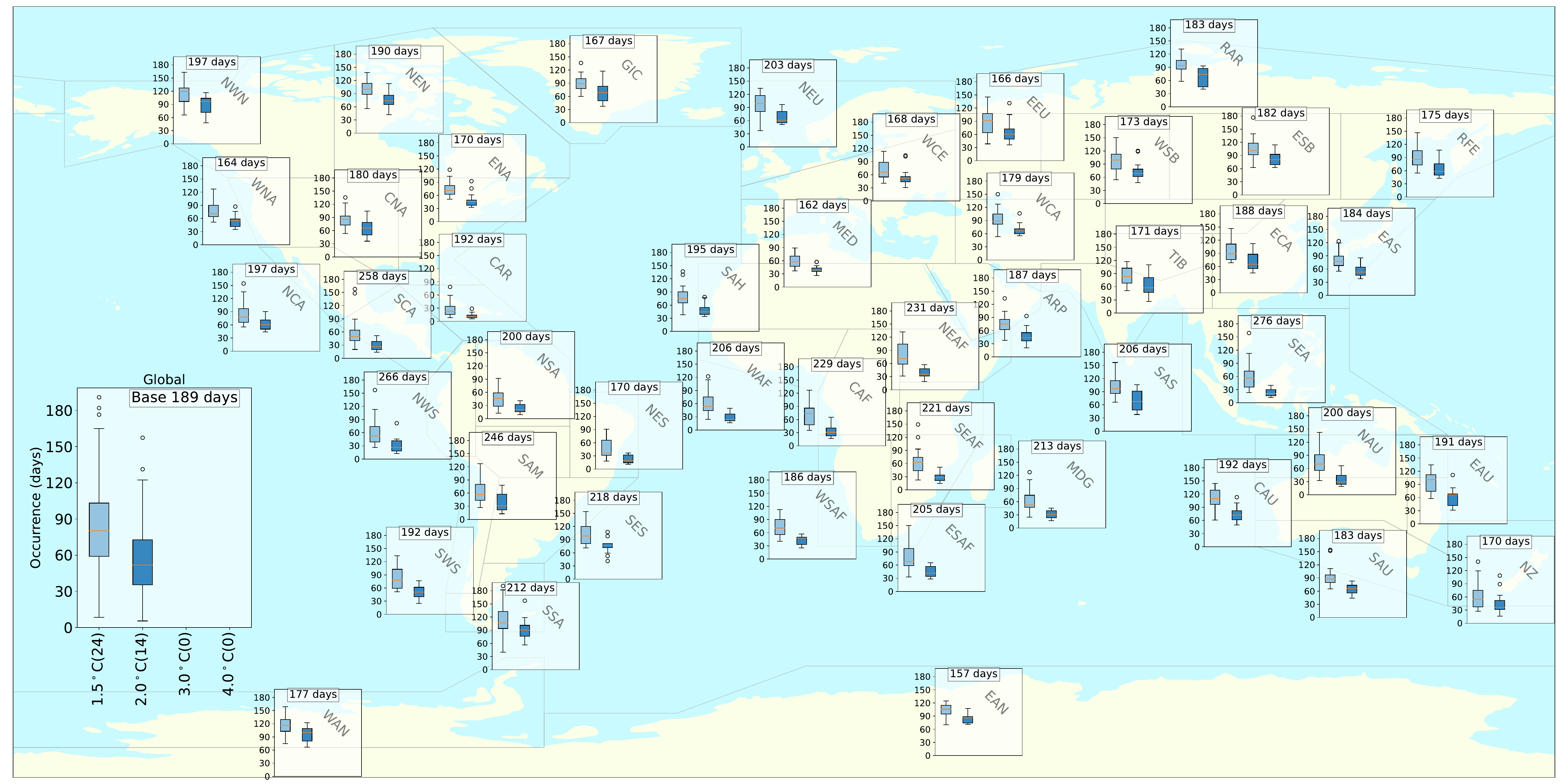}
    \caption{Multi-model median of return period for 1-year temperature events compared to the 1980-2010 period under GWL 1.5, 2, 3 and 4$^\circ$C relative to 1850-1900 baseline for SSP1-2.6 scenario. Blue, green, pink and red boxes represent 1.5$^\circ$C, 2.0$^\circ$C, 3.0$^\circ$C and 4.0$^\circ$C, respectively. The orange lines inside the boxes show CMIP6 multi-model median, and the boxes extend between the first quartile (Q1) to the third quartile (Q3) of the data, i.e. inter-quartile range (IQR). The vertical lines, whiskers, stretch out 1.5 IQR from the box. The circles represent the model outliers. The length of the hot period used for return period calculations, i.e. number of days in a 10-year period, is shown on top right corner of each plot. The global return periods are shown on the left. The reason for more outlier points in the global box plot is the differences between regional return periods.}
    \label{fig:1yearmap126}
\end{figure}

\begin{figure}[ht]
    \centering
    \includegraphics[width=\textwidth]{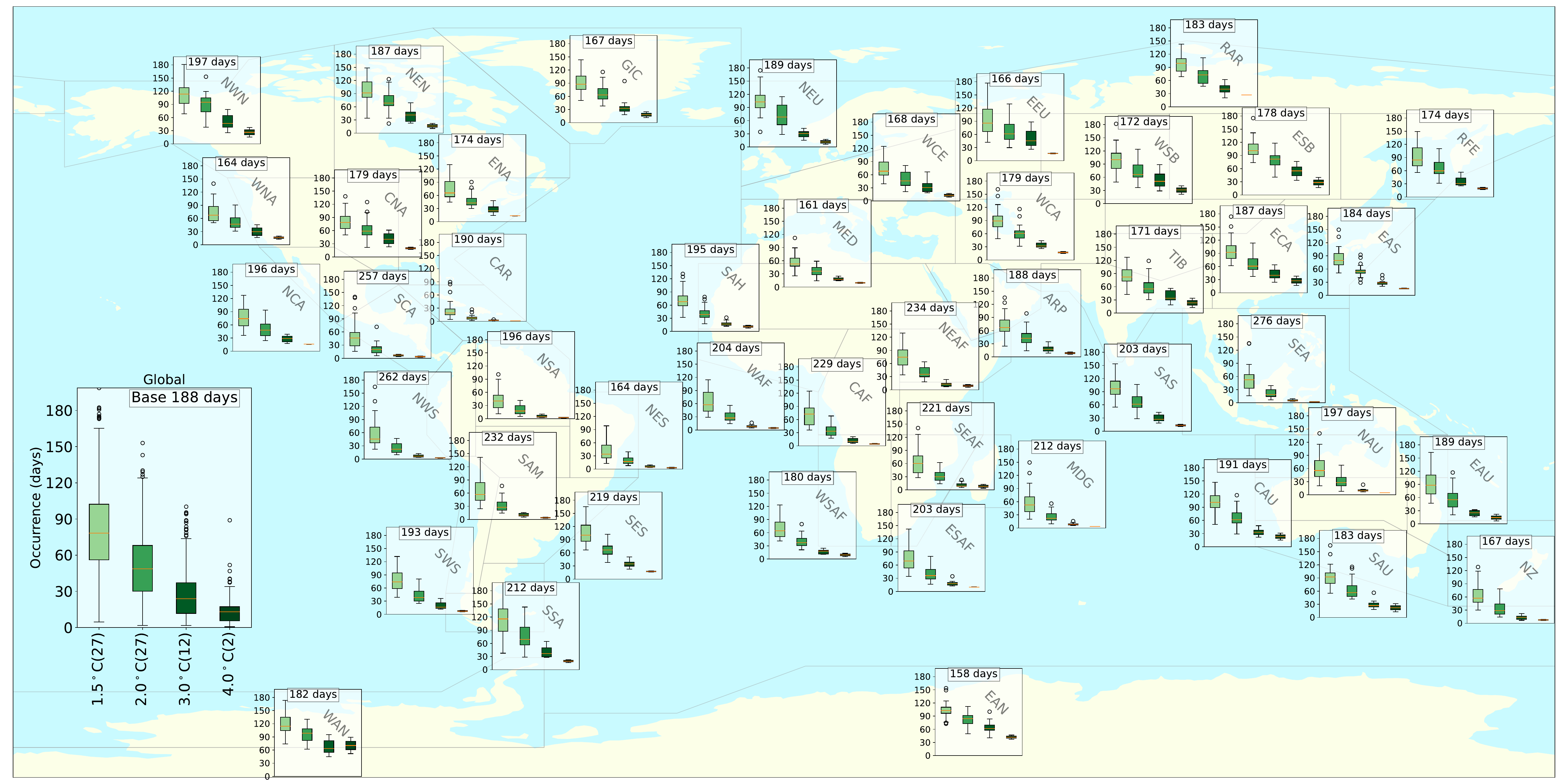}
    \caption{Same as Figure \ref{fig:1yearmap126} but for SSP2-4.5}
    \label{fig:1yearmap245}
\end{figure}

\begin{figure}[ht]
    \centering
    \includegraphics[width=\textwidth]{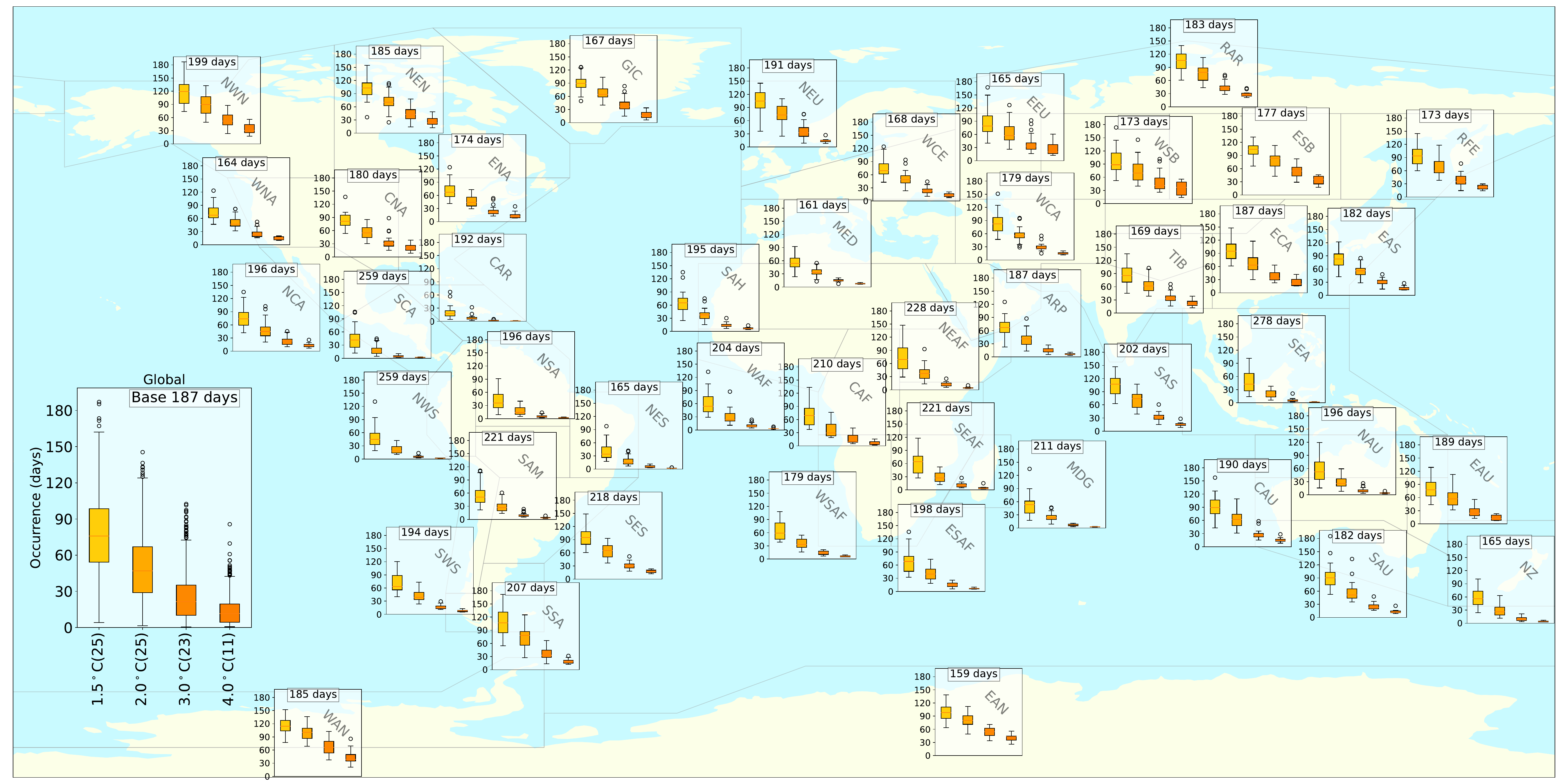}
    \caption{Same as Figure \ref{fig:1yearmap126} but for SSP3-7.0}
    \label{fig:1yearmap585}
\end{figure}

\begin{figure}[ht]
    \centering
    \includegraphics[width=\textwidth]{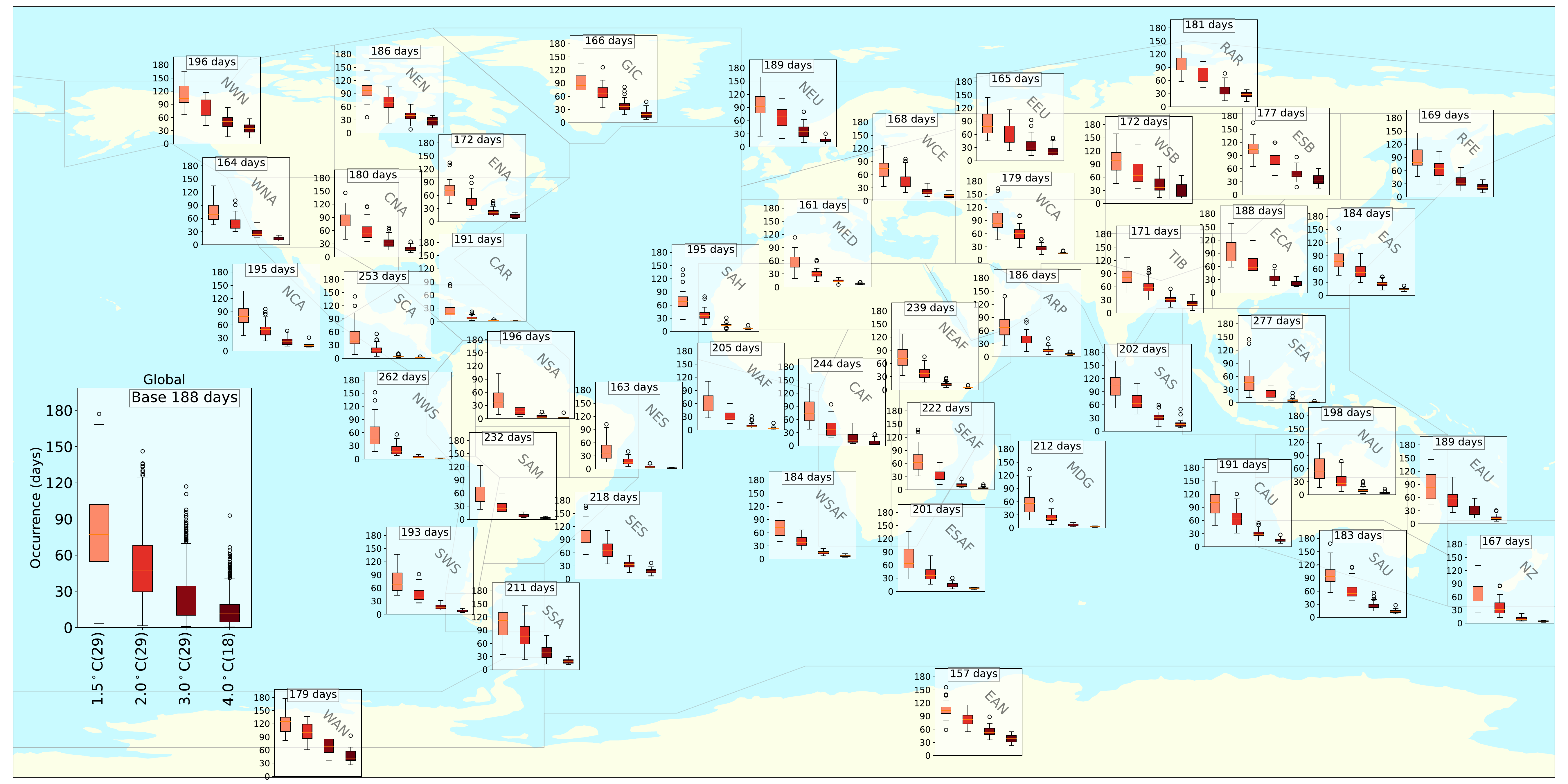}
    \caption{Same as Figure \ref{fig:1yearmap126} but for SSP5-8.5}
    \label{fig:1yearmap370}
\end{figure}

\begin{figure}[ht]
    \centering
    \includegraphics[width=\textwidth]{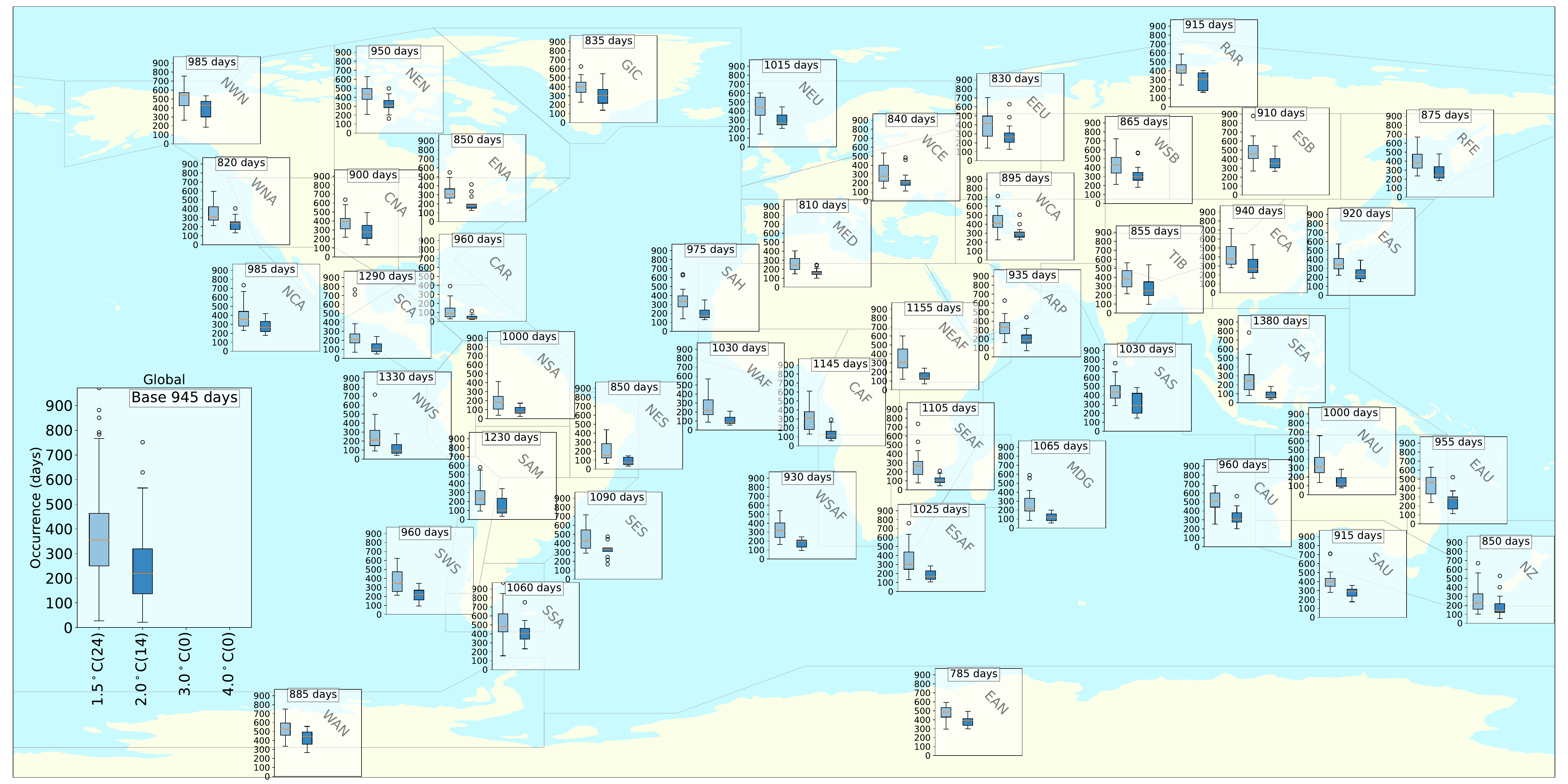}
    \caption{Multi-model median of return period for 5-year temperature events under SSP1-2.6}
    \label{fig:5yearmap126}
\end{figure}

\begin{figure}[ht]
    \centering
    \includegraphics[width=\textwidth]{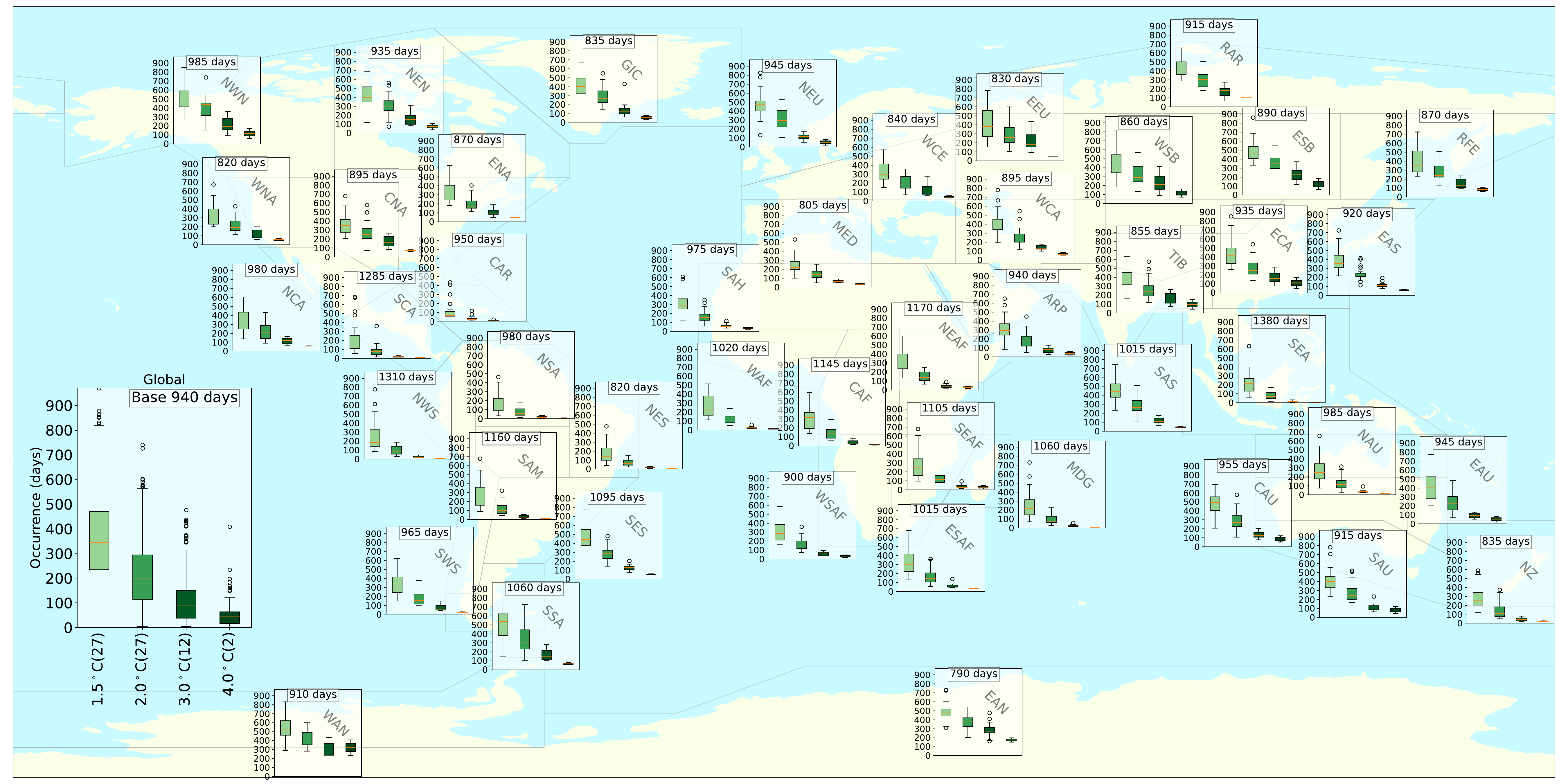}
    \caption{Same as Figure \ref{fig:5yearmap126} but for SSP2-4.5}
    \label{fig:5yearmap245}
\end{figure}

\begin{figure}[ht]
    \centering
    \includegraphics[width=\textwidth]{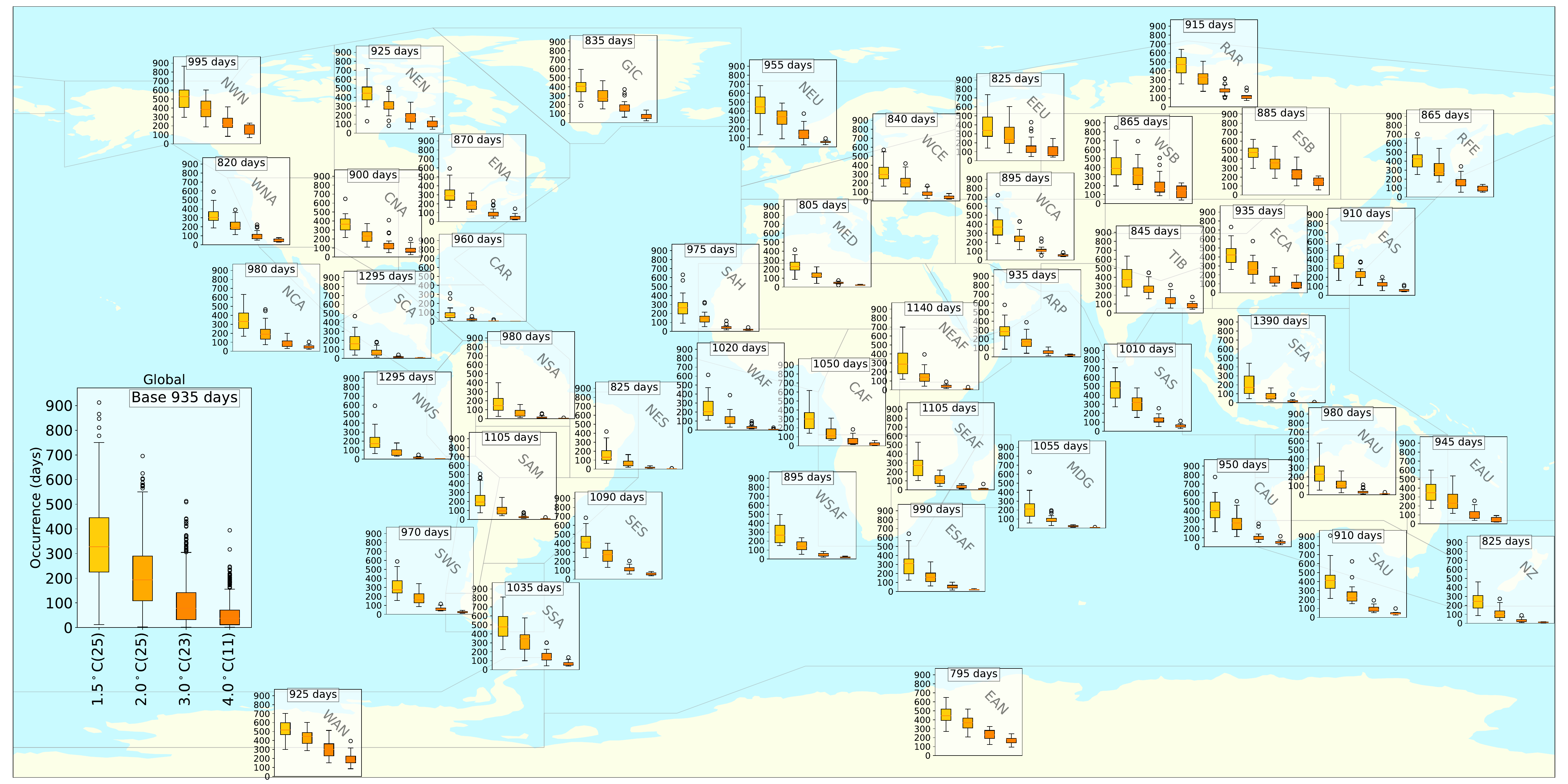}
    \caption{Same as Figure \ref{fig:5yearmap126} but for SSP3-7.0}
    \label{fig:5yearmap370}
\end{figure}

\begin{figure}[ht]
    \centering
    \includegraphics[width=\textwidth]{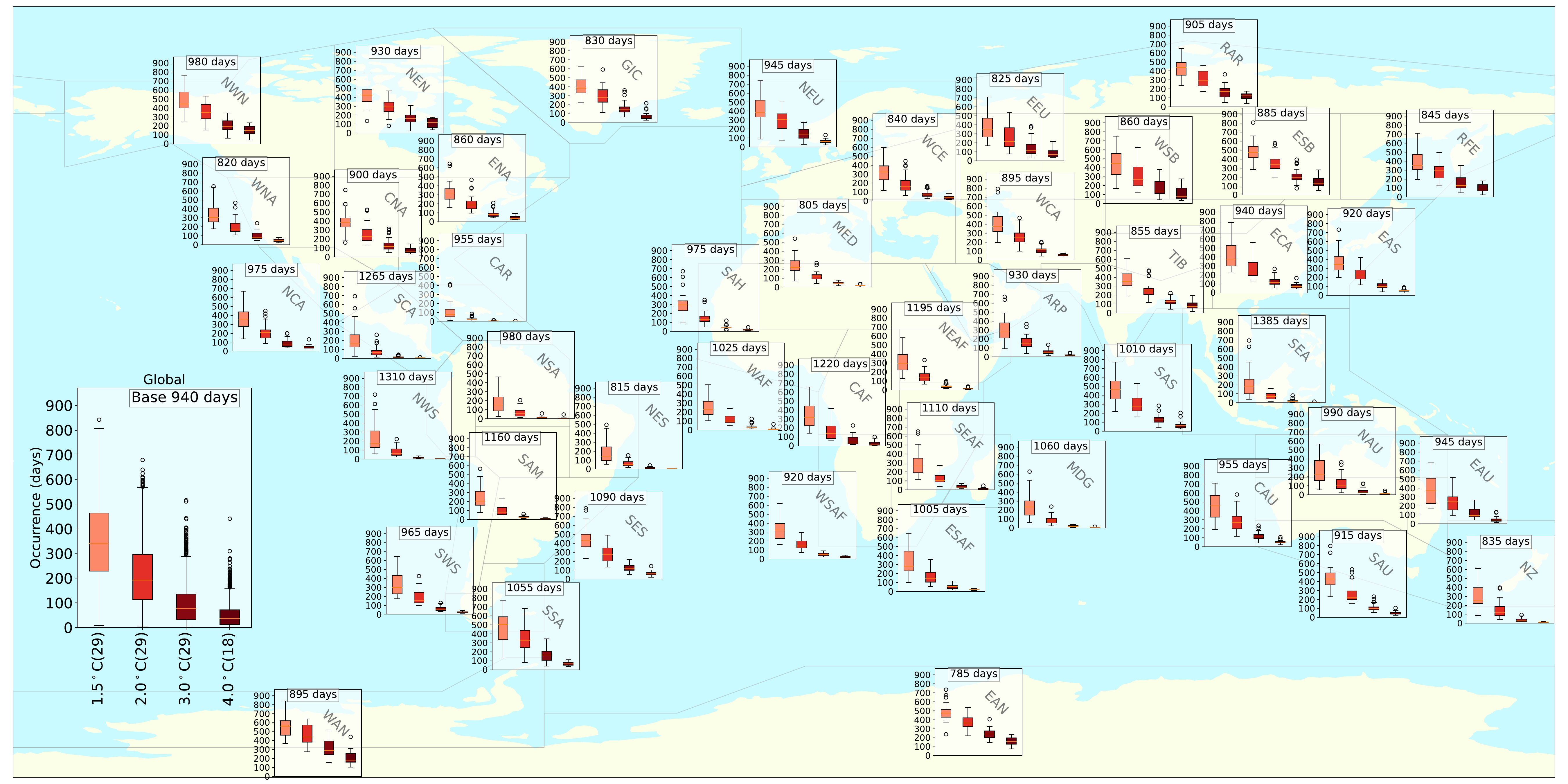}
    \caption{Same as Figure \ref{fig:5yearmap126} but for SSP5-8.5}
    \label{fig:5yearmap585}
\end{figure}

\begin{figure}[ht]
    \centering
    \includegraphics[width=\textwidth]{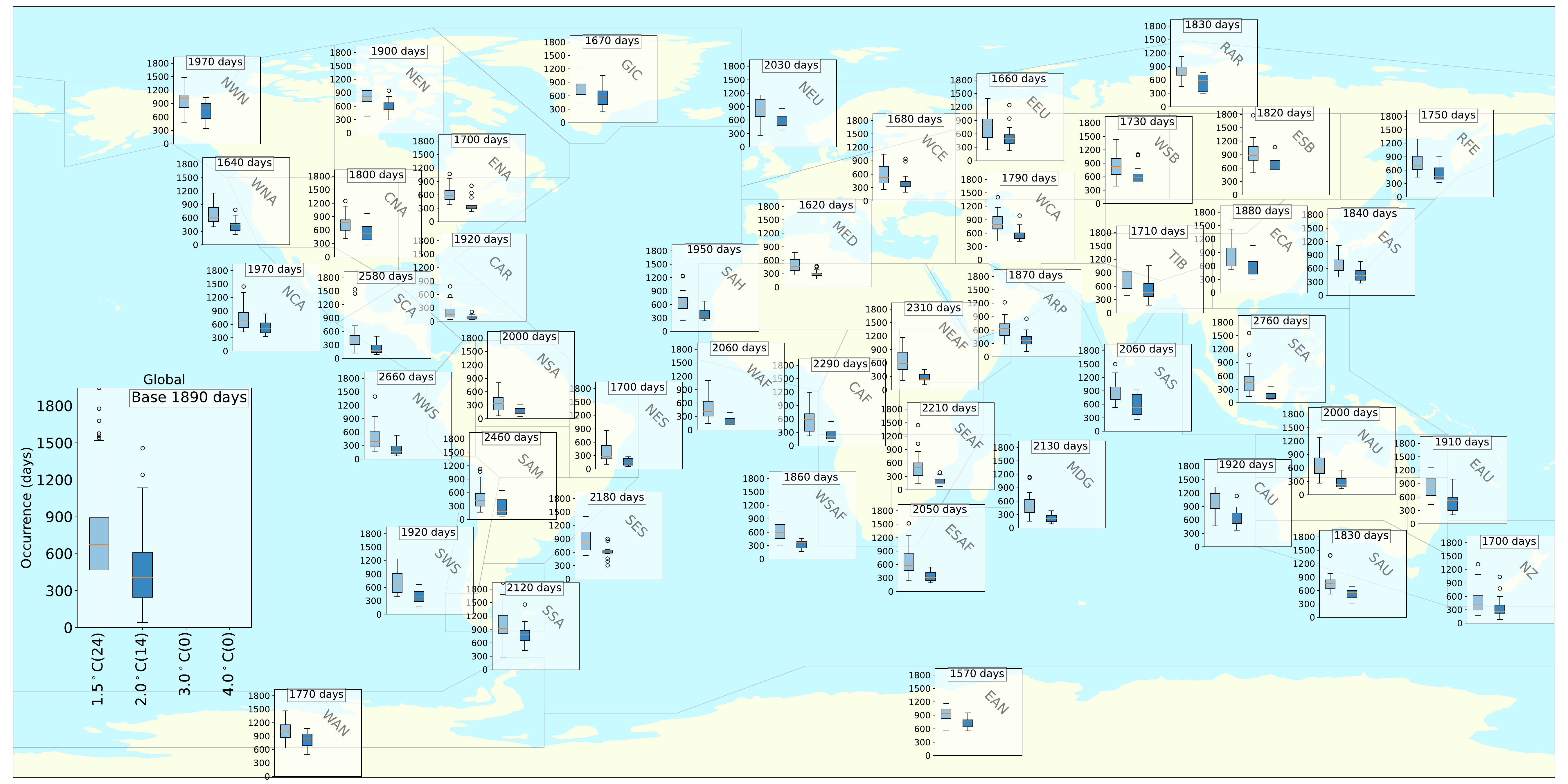}
    \caption{Multi-model median of return period for 10-year temperature events under SSP1-2.6}
    \label{fig:10yearmap126}
\end{figure}

\begin{figure}[ht]
    \centering
    \includegraphics[width=\textwidth]{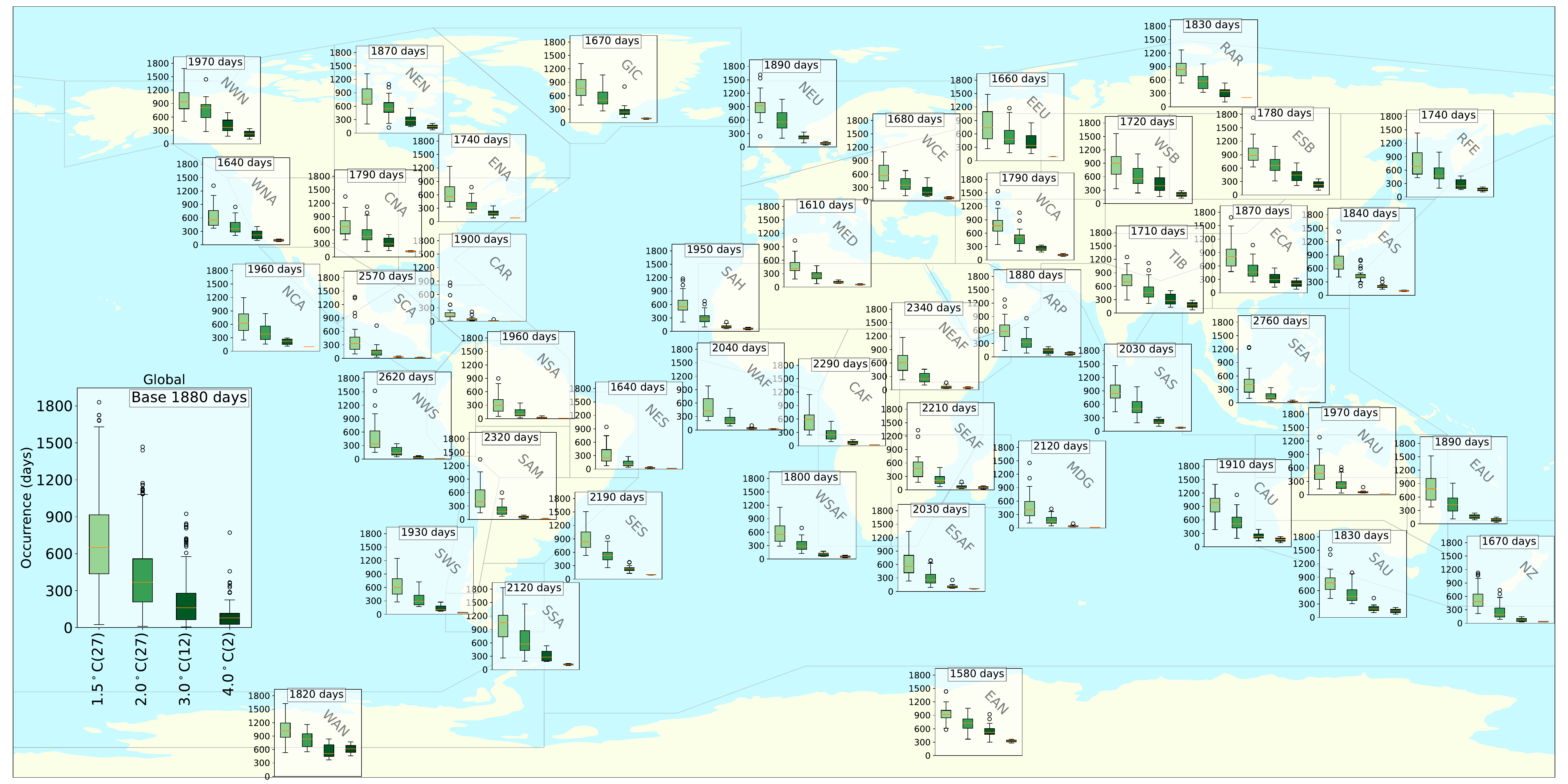}
    \caption{Same as Figure \ref{fig:10yearmap126} but for SSP2-4.5}
    \label{fig:10yearmap245}
\end{figure}

\begin{figure}[ht]
    \centering
    \includegraphics[width=\textwidth]{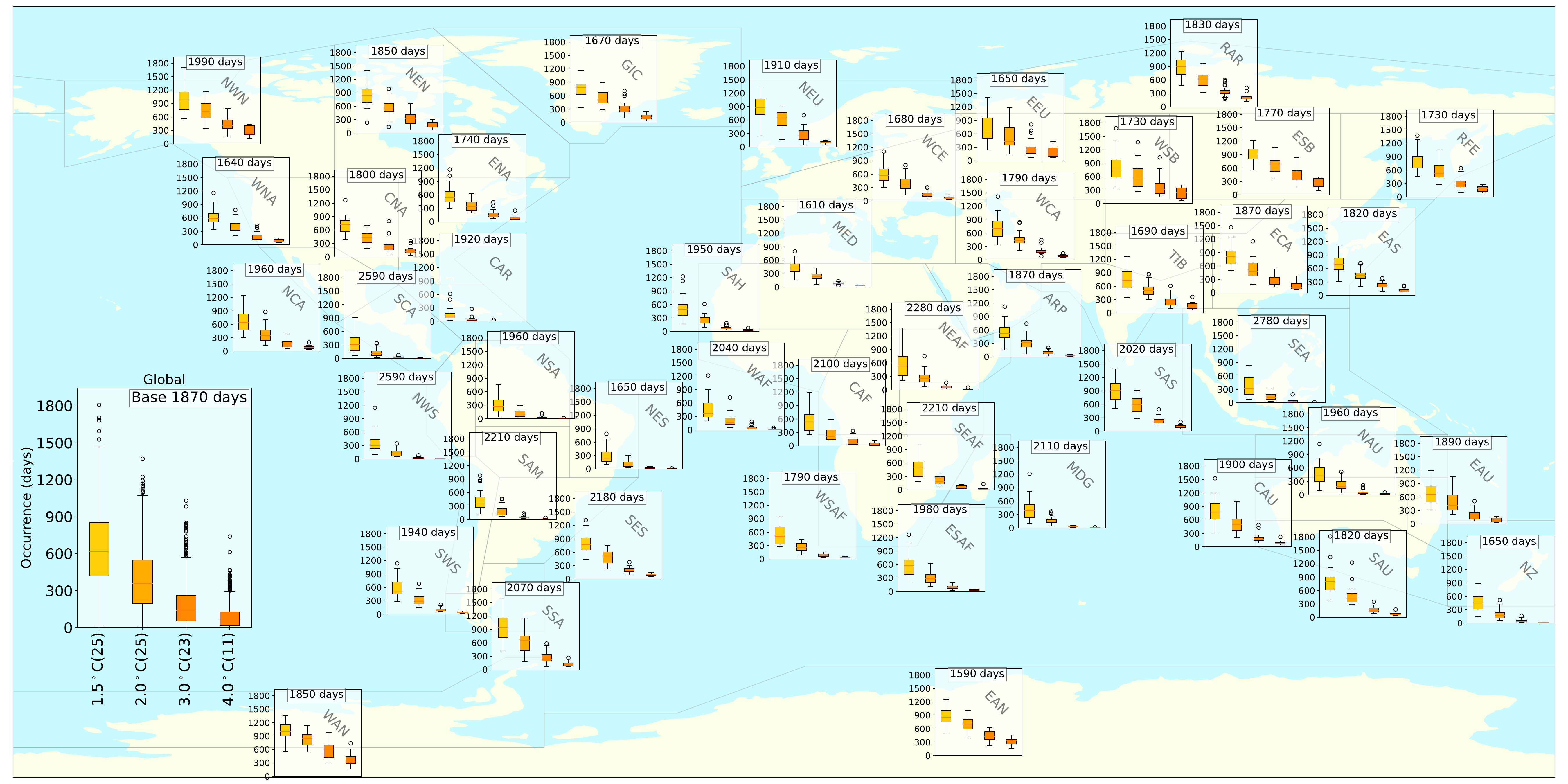}
    \caption{Same as Figure \ref{fig:10yearmap126} but for SSP3-7.0}
    \label{fig:10yearmap370}
\end{figure}

\begin{figure}[ht]
    \centering
    \includegraphics[width=\textwidth]{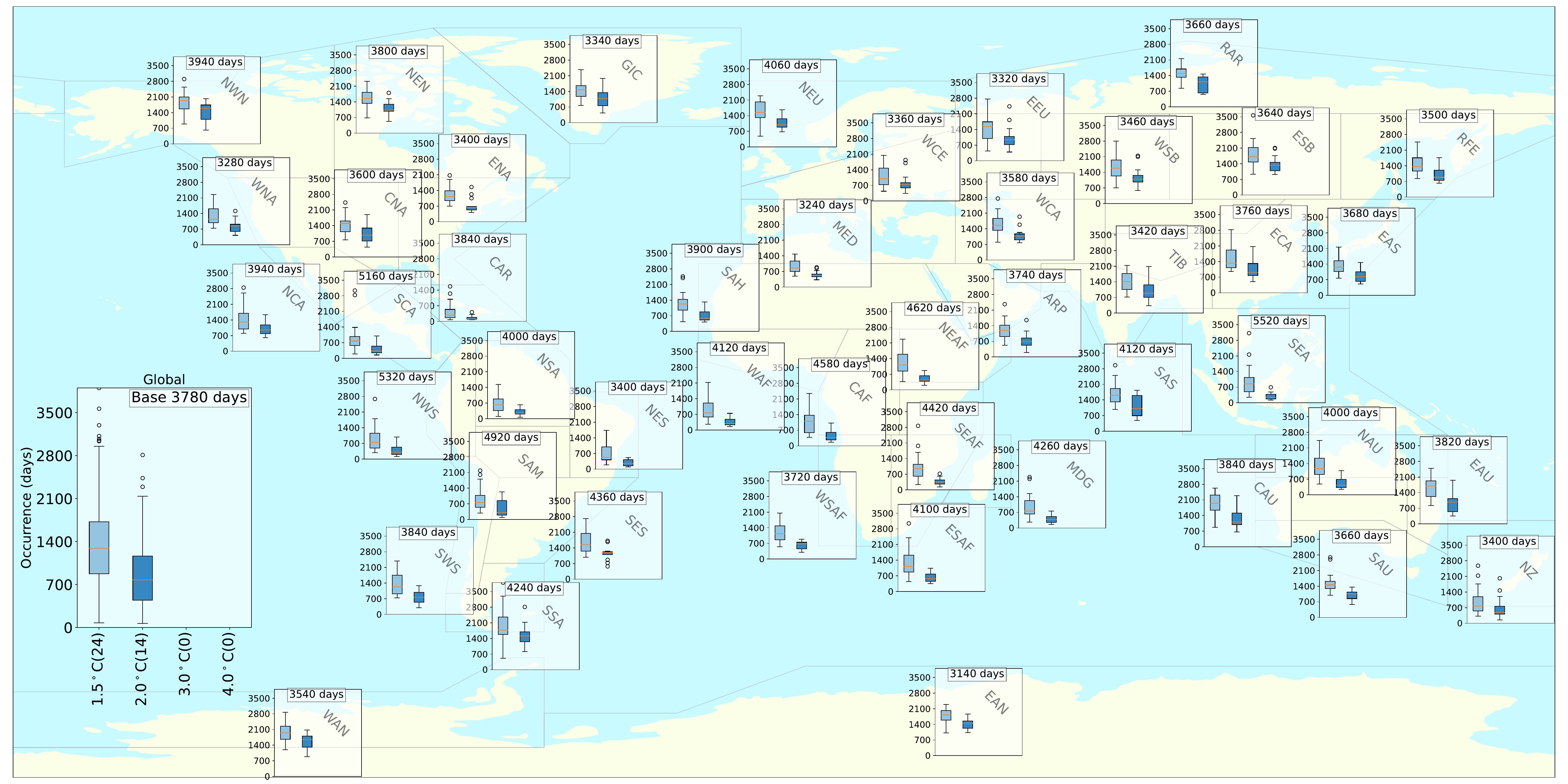}
    \caption{Multi-model median of return period for 20-year temperature events under SSP1-2.6}
    \label{fig:20yearmap126}
\end{figure}

\begin{figure}[ht]
    \centering
    \includegraphics[width=\textwidth]{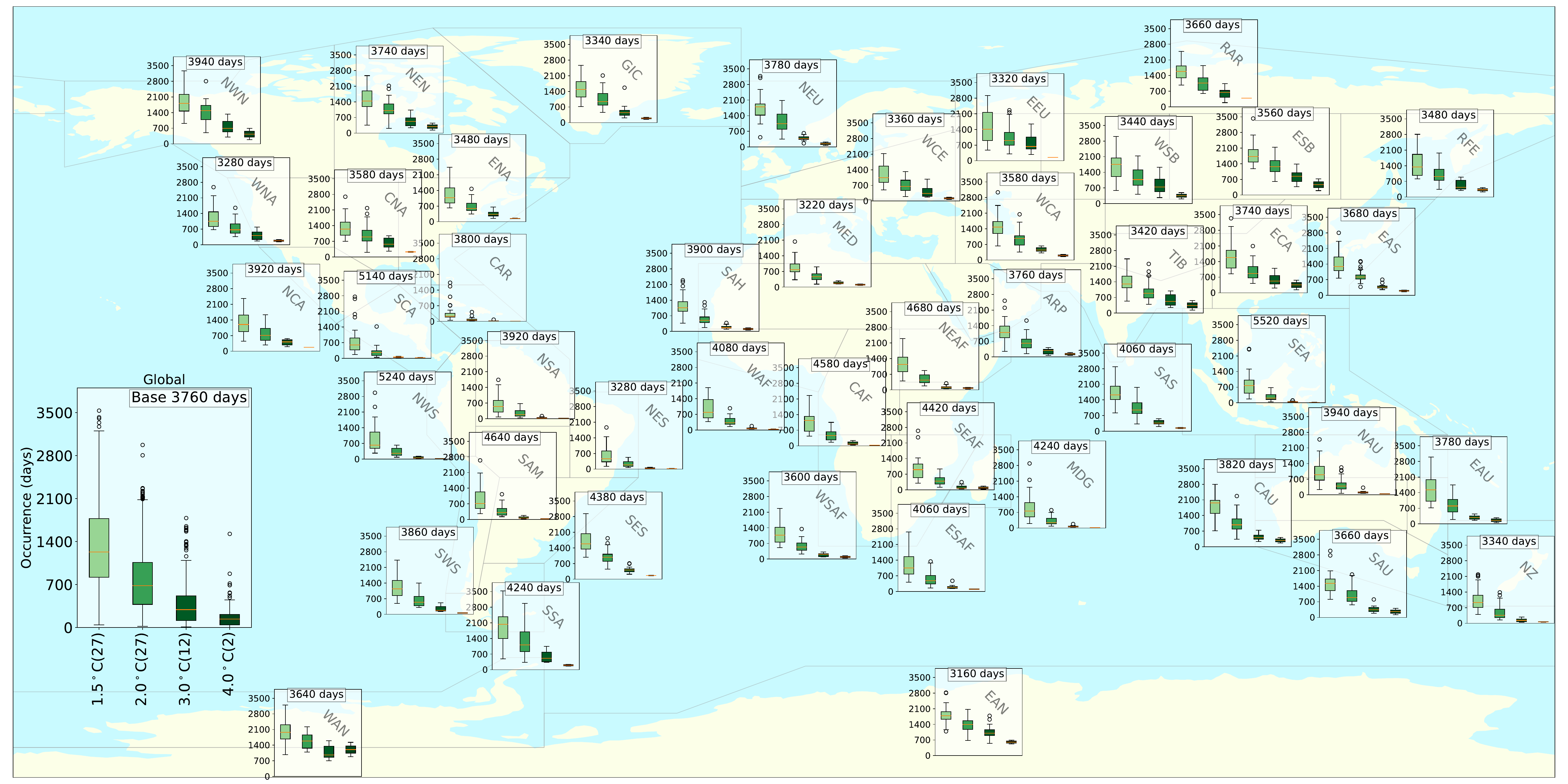}
    \caption{Same as Figure \ref{fig:20yearmap126} but for SSP2-4.5}
    \label{fig:20yearmap245}
\end{figure}

\begin{figure}[ht]
    \centering
    \includegraphics[width=\textwidth]{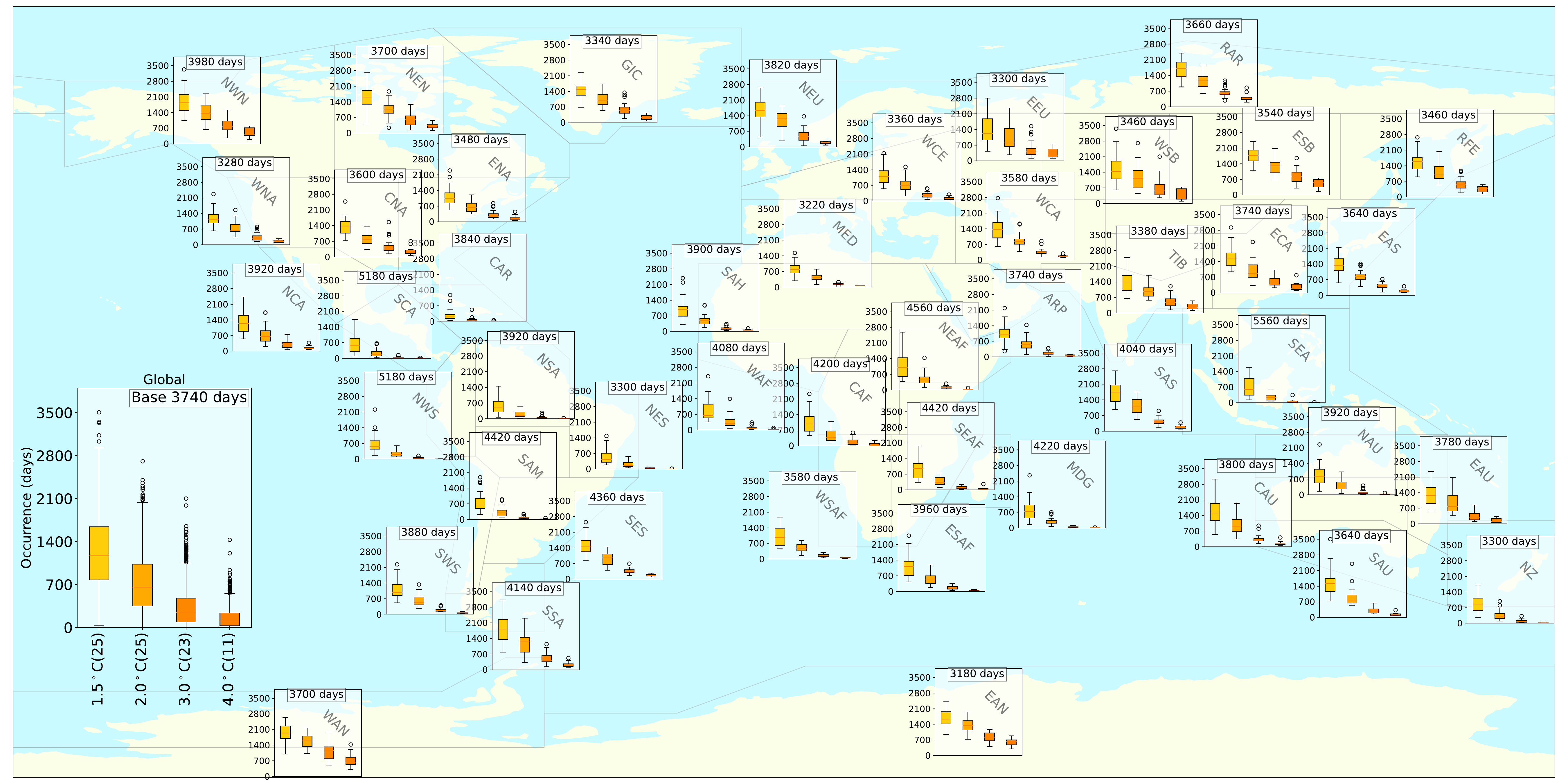}
    \caption{Same as Figure \ref{fig:20yearmap126} but for SSP3-7.0}
    \label{fig:20yearmap370}
\end{figure}

\begin{figure}[ht]
    \centering
    \includegraphics[width=\textwidth]{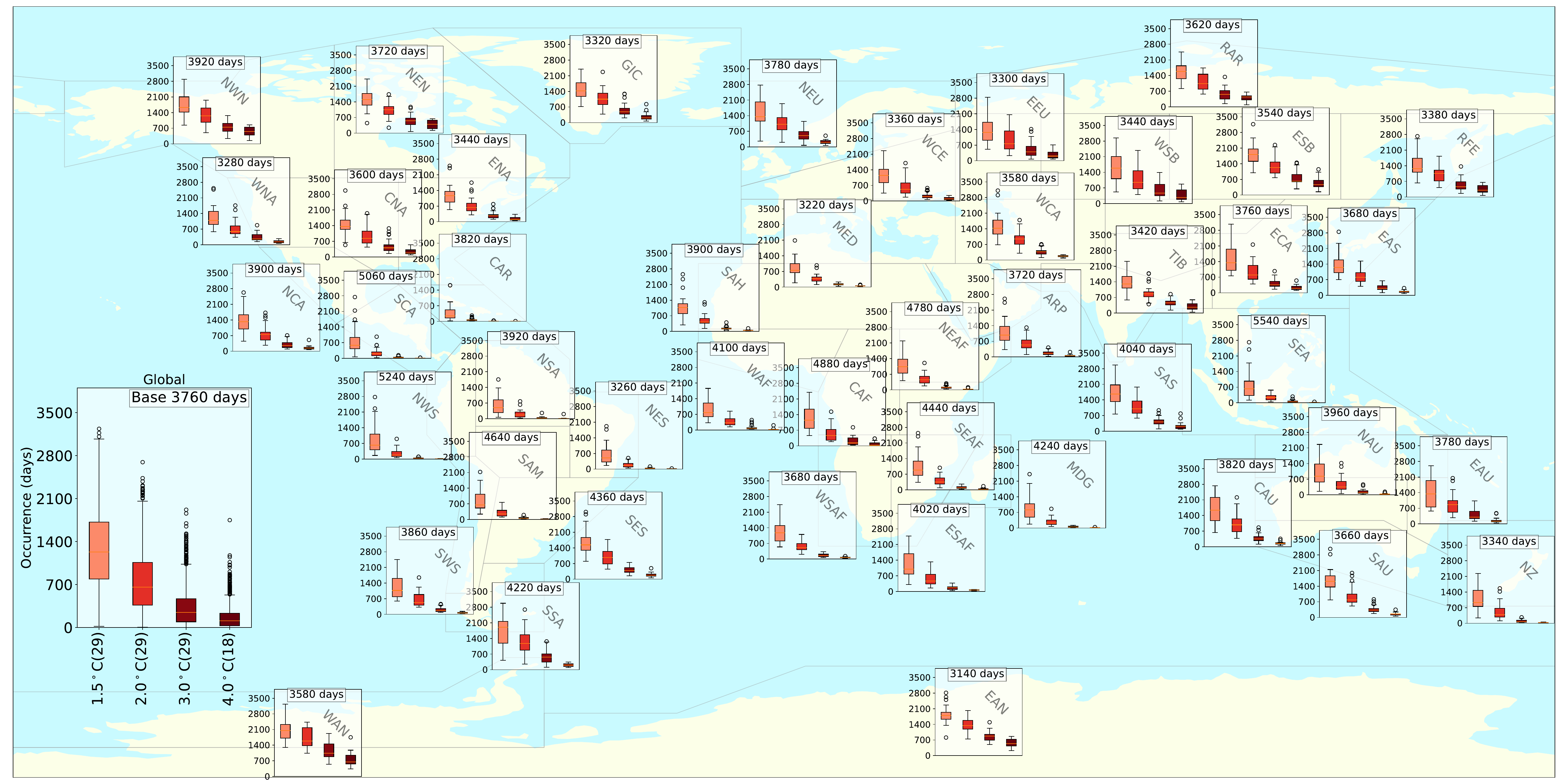}
    \caption{Same as Figure \ref{fig:20yearmap126} but for SSP5-8.5}
    \label{fig:20yearmap585}
\end{figure}

\noindent\textbf{Contents of this file}
\begin{enumerate}
\item Text S\ref{sec:gft} to S\ref{sec:gmm}
\item Table \ref{tab:KStest}  to \ref{tab:gmm_periods}
\item Figures \ref{fig:gmm_comparison} to \ref{fig:20yearmap585}
\end{enumerate}
\textbf{Introduction} This supplementary information includes more detailed information about testing different distributions on daily maximum temperature data and additional figures for other global warming levels and SSP scenarios. Future periods of datasets for different GWL levels are also presented. 

\section{Goodness-of-fit tests}
\label{sec:gft}
To demonstrate the goodness-of-fit for Gaussian Mixture Models (GMM), here we used 31-year historical daily maximum temperature data. The daily maximum temperature data is exported from a grid cell located at East 28.9784 and West 41.0082 coordinates (near Istanbul, Turkey) from MPI-ESM1-2-HR climate simulation using the nearest neighbour method. The apparent bimodality of daily maximum temperature data can be seen in Figure \ref{fig:gmm_comparison}. We used Kolmogorov-Smirnov statistics to test the goodness-of-fit for different unimodal distributions; normal and GEV distributions, and GMM distribution. We used GEV distribution with the same location and scale parameters but different shape parameters, $\xi$. The shape parameters of 0, 0.5, and -0.5 are known as the Gumbel, Fréchet and Weibull distributions, respectively. We also used a GEV distribution with its shape parameter estimated from the data. The KS-test returns the test statistic and the p-value. The null hypothesis is that the two samples are drawn from the same distribution. The smaller the p-value, the more likely it is that the two samples are drawn from different distributions. All distributions except GMM have a p-value smaller than 0.05 significance level, so they are 'statistically different' from the raw data. However, the GMM has a p-value of 0.73, so it is not statistically different from the data. KS-test scores for different distributions are shown in Table \ref{tab:KStest}.

\section{Choosing the best number of components}
\label{sec:gmm}
GMM use the Expectation-Maximization algorithm to fit Gaussian components to data. Bayesian Information Criterion (BIC) scores are used to assess the best number of components for the mixture model. Different numbers of components are scored, and the number of components with the lowest BIC score is kept as the best fit. However, we observed that the BIC scores are very similar for more than three components in some grid cells. 
Figure \ref{fig:gmm_selection} shows the distribution of raw temperatures in the top plot. Here, we tested a grid cell (Lat=43.48, Lon=19.69) in the MED region from the MPI-ESM1-2-HR model for the historical period of 1985-2014 and the future period of SSP5-8.5 GWL3$^\circ$C. Predicted GMM for historical and future periods are plotted with green and red lines. The plots in the middle row of the figure display the initial choices of GMM components for the historical and future periods based on the lowest BIC scores. In the case of this grid cell, a GMM with 7 Gaussian components was determined to be the best fit for the historical period, while a GMM with 2 Gaussian components was the best fit for the future period. The initial choices are shown with asterisks in the plots in the middle row of the figure. The plots in the bottom row of the figure show the best fit for the historical and future periods after applying the highest gradient change and unimodality tests. For the historical period, 2 Gaussian components indeed are a better fit. We would like to point out that, our tests validate the initial choice of the lowest BIC score for the future period.

\section{Origins of bimodality}
\begin{figure}
    \centering
    \includegraphics[width=0.9\textwidth]{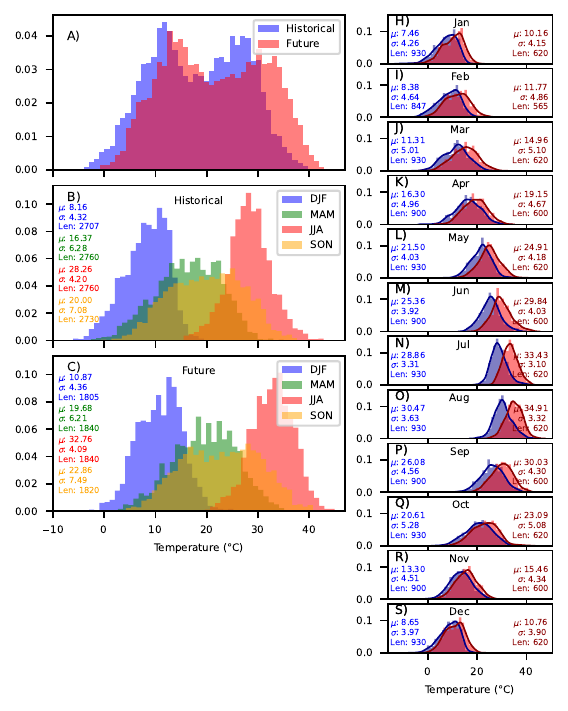}
    \caption{A) Blue and red bars represent multi-year raw daily maximum temperature data from a grid cell (Lat=43.48, Lon=19.69) in the MED region from the MPI-ESM1-2-HR model for the 30-year historical period of 1985-2014 and SSP5-8.5 GWL3$^\circ$C 20-year future period. B-C) Seasonal distributions of temperatures for the historical and future periods. The parameters for each distribution are listed on the left side with matching colours. H-S) The monthly raw daily maximum temperature data is represented by the blue and red distributions.}
    \label{fig:gmm_months_analyse}
\end{figure}

To understand the underlying factors of the bimodal pattern, temperature distributions were analyzed from grid cells with distinct bimodality across various timeframes, including different months and seasons as shown in Figure \ref{fig:gmm_months_analyse}. The analysis revealed that the bimodal distribution was primarily influenced by the winter and summer seasons. These seasons appeared to be the main contributors to the peaks observed in the bimodal distribution. The transition from winter to summer happened rapidly, leading to a more pronounced separation of the temperature modes. As a result, the distributions during transitional seasons, such as spring and autumn, were wider, covering a broader range of temperature values compared to the more distinct distributions observed during winter and summer, which covered a very small range of temperature values.

\bibliography{bibliography.bib}
\end{article}
\clearpage

\begin{table}[ht]
    \centering
    \caption{KS test results for different distributions.}
    \label{tab:KStest}
    \begin{tabular}{lcr}
    Distribution  &         D &        \textit{p}-value \\ \hline
    Normal Distribution            &  0.962766 &  $\ll$ .001  \\
    GEV Distribution ($\xi$=0.5)   &  0.182460 &  $\ll$ .001 \\
    GEV Distribution ($\xi$=0)     &  0.102005 &  $\ll$ .001  \\
    GEV Distribution ($\xi$=-0.5)  &  0.091142 &  $\ll$ .001 \\
    GEV Distribution ($\xi$=-0.33) &  0.054491 &  $\ll$ .001  \\
    GMM Distribution               &  0.009076 &  0.73  \\ \hline
    \end{tabular}
\end{table}
\begin{table}
\label{tab:regions}
\caption{46 IPCC land and land-ocean regions used in the study \cite{Iturbide2020} (regionDescription.xlsx).}
\end{table}

\begin{table}[]
    \centering
    \caption{The start and end years of 20-year Global Warming Level periods for CMIP6 dataset under SSP scenarios. The start and end years of these 20-year GWL periods for each CMIP6 simulation are obtained from \cite{gwl_periods} (CMIP6\_SSP\_GWL\_periods.xlsx)}
    \label{tab:gmm_periods}
\end{table}

